\documentclass{article}

\usepackage{arxiv}

\usepackage[utf8]{inputenc} 
\usepackage[T1]{fontenc}    
\usepackage{url}            
\usepackage{booktabs}       
\usepackage{amsfonts}       
\usepackage{nicefrac}       
\usepackage{microtype}      
\usepackage{lipsum}		
\usepackage{graphicx}
\usepackage{doi}

\usepackage[T1]{fontenc}
\usepackage{amsmath}
\usepackage{amsfonts}
\usepackage{amssymb}
\usepackage[justification=centering]{caption}
\usepackage{subcaption}
\usepackage{blindtext}
\usepackage{algorithm}
\usepackage{algpseudocode}
\usepackage[dvipsnames]{xcolor}
\usepackage{comment}
\usepackage{float}
\usepackage{xcolor}
\usepackage{pgfplots}
\usepackage{xparse}
\usepackage{wrapfig}
\usepackage{todonotes}
\usepackage{multirow}
\usepackage{caption}
\usepackage{lipsum}
\usepackage{booktabs}
\usepackage{stmaryrd}
\usepackage{listings}

\newcommand{\etal}{\textit{et al.}}
\DeclareMathOperator*{\argmax}{arg\,max}

\lstdefinestyle{CStyle}{
	keywordstyle=\color{magenta},
	language=C
}

\floatname{algorithm}{Algorithm}

\title{Like an Open Book? Read Neural Network Architecture with Simple Power Analysis on 32-bit Microcontrollers}

\date{}

\author{ {Raphaël Joud, Pierre-Alain Moëllic, Simon Pontié} \\
	CEA Tech, Centre CMP, Equipe Commune CEA Tech - Mines Saint-Etienne, F-13541 Gardanne, France \\
	Univ. Grenoble Alpes, CEA, Leti, F-38000 Grenoble, France\\
	\texttt{\{name\}.\{surname\}@cea.fr} \\
	\And
	{Jean-Baptiste Rigaud} \\
	Mines Saint-Etienne, CEA, Leti, Centre CMP,\\ F-13541 Gardanne France\\
	\texttt{rigaud@emse.fr} \\
}

\renewcommand{\headeright}{(arxiv version)}
\renewcommand{\undertitle}{(arxiv version - Accepted CARDIS 2023)}
\renewcommand{\shorttitle}{Like an Open Book? Read DNN Architecture with SPA on 32-bit MCU}

\begin{document}
\maketitle

\begin{abstract}
Model extraction is a growing concern for the security of AI systems. For deep neural network models, the architecture is the most important information an adversary aims to recover.     
 
Being a sequence of repeated computation blocks, neural network models deployed on edge-devices will generate distinctive side-channel leakages. The latter can be exploited to extract critical information when targeted platforms are physically accessible.
By combining theoretical knowledge about deep learning practices and analysis of a widespread implementation library (ARM CMSIS-NN), our purpose is to answer this critical question: \textit{how far can we extract architecture information by simply examining an EM side-channel trace?}
For the first time, we propose an extraction methodology for traditional MLP and CNN models running on a high-end 32-bit microcontroller (Cortex-M7) that relies only on simple pattern recognition analysis. Despite few challenging cases, we claim that, contrary to parameters extraction, the complexity of the attack is relatively low and we highlight the urgent need for practicable protections that could fit the strong memory and latency requirements of such platforms.
\end{abstract}

\section{Introduction}
\label{intro}

Deployment of Deep Neural Network (DNN) models continues to gain momentum, typically with Internet of Things (IoT) applications with microcontroller-based platforms. However, their security is regularly challenged with works focused on availability, integrity and confidentiality threats~\cite{papernot2018sok}. Latter topic keeps gathering growing attention from the adversarial Machine Learning (ML) community with \textit{Model Extraction}~\cite{carlini2020cryptanalytic,jagielski2020high,joud2023practical} becoming a major concern.

An attacker performs a model extraction attack either to steal a well-trained model performance or to precisely recover its characteristics to obtain a \textit{clone model}. Additionally, it would allow the adversary to enhance his level of control over the victim system to design more adapted and powerful attacks. Therefore, model extraction attacks have been the subject of a growing number of works in recent years with different adversarial scenarios regarding the level of knowledge related to the model and the training data distribution. In the literature, model architecture and parameters are usually studied separately. Many efforts have been brought to parameters recovery with milestones works relying on the assumption that model architecture is known by the attacker~\cite{carlini2020cryptanalytic,gongye2020reverse,jagielski2020high,maji2021leaky,rakin2022deepsteal}. However, such an assumption is quite strong and can be questioned as architecture details are generally not disclosed. However, whatever the attacker's objectives, knowledge of victim model architecture significantly increases adversarial ability. 

This work is focused on architecture extraction of models embedded on 32-bit microcontrollers thanks to ARM CMSIS-NN library. Our purpose is to know how far an adversary can extract information from a victim model architecture by jointly exploiting the knowledge of the deployment library and very limited side-channel (EM) traces (averaged trace from a single input). Surprisingly and worryingly, we show that (with methods and classical ML expertise) almost all the most important information and hyper-parameters are reachable because of the high \textit{repetitiveness} of the underlying computations. More particularly, for convolutional neural networks, we demonstrate a \textit{Russian doll} effect: one regular EM pattern is related to one hyper-parameter and zooming in presents other patterns related to others hyper-parameters, and so on.    
%
\\

\noindent\textbf{Illustrations, code and data availability.} This work relies on the visual analysis of side-channel traces with many (colored) illustrations to explain our approach. For conciseness purpose, we select the most representative ones. We propose additional contents in a public repository\footnote{\url{https://gitlab.emse.fr/securityml/model-architecture-extraction}} with codes and data (traces).       

\section{Background}
\label{bckgrnd}

\subsection{Neural network models}
\label{bckgrnd:NN}

\noindent We consider a supervised deep neural network $M_W$, with $W$ its internal parameters. When fed with an input $x \in \mathbb{R}^d$, the model outputs a set of \textit{prediction scores} $M_W(x) \in \mathbb{R}^{K}$ with $K$, the number of labels. Then, the output predicted label is $\hat{y} = \argmax(M_W(x))$. 
 
We note $\mathcal{A}_M$ the architecture of the model $M$ that corresponds to the organisation of its \textit{layers}. $\mathcal{A}_M$ is defined by the nature of each layer, their connections, size and hyper-parameters (non-trainable ones, e.g. the number of convolutional kernels). We note $L$ the number of layers. In this work, we only consider feedforward models with layers stacked horizontally. 
\\

\noindent\textbf{MultiLayer Perceptron (MLP)} are composed of several vertically stacked \textit{neurons} (or \textit{perceptron}) called \textit{dense layers} (also named \textit{fully-connected} or even \textit{linear}). Each perceptron first processes a weighted sum of its trainable parameters $w$ and $b$ (called  \textit{bias}) with the input $x=(x_0,..., x_{n-1}) \in \mathbb{R}^n$. Then, it non-linearly maps the output thanks to an \textit{activation function} $\sigma$: $a(x)=\sigma(w_{0}x_0+...+w_{n-1}x_{n-1}+b)$, where $a$ is the perceptron output. For MLP, a neuron from layer $l$ gets inputs from all neurons belonging to previous layer $l-1$. 
\\

\noindent\textbf{Convolutional Neural Network (CNN)} process input data with a set of convolutional \textit{kernels} (also called \textit{filters}). The kernels are usually low-dimensional squared matrices (e.g. $3\times3$ for image classification). In addition, kernels third dimension matches the number of channels of the input tensor $C_{in}$ (e.g. for RGB images, $C_{in}=3$). As such, for a \textit{convolutional layer} (hereafter shorted to conv. layer) composed of $K$ kernels of size $Z \times Z \times C_{in}$ applied to an input tensor of size $H_{in}\times H_{in} \times C_{in}$ (we use square inputs for simplicity), the weight tensor $W$ will have the shape $[K,Z,Z,C_{in}]$ (i.e. $K\cdot C_{in}\cdot Z^2$ parameters without bias, $(K+1)\cdot C_{in}\cdot Z^2$ otherwise). Additionally, The number of output tensor channels is $C_{out}=K$. Eq.~\ref{eq_cnn} is a convolution (without bias) expressed as dot-products between kernel and local regions of the input tensor, classically processed by a sliding window. With $Y$ the output tensor, we have $\forall k,l,n \in \llbracket0,H_{in}\llbracket^{2} \times \llbracket0,C_{out}\llbracket$:

\begin{equation}
Y_{k,l,n}=\sigma\Big(\sum_{m=0}^{C_{in}-1}\sum_{i=0}^{Z-1}\sum_{j=0}^{Z-1}W_{i,j,m,n}\cdot X_{k+i,l+j,m}\Big) 
\label{eq_cnn}
\end{equation}

\noindent Other hyper-parameters are \textit{Padding} $P$ and \textit{Stride} $S$ that, respectively, adds extra dimensions ($P$) to handle the borders of the input tensor\footnote{With $P=0$, borders are not considered and the output tensor is shorter} and enables to downsample (by $S$) the sliding\footnote{$S=1$ is the standard default sliding, one element at a time.}. The output tensor shape is defined as in Eq.~\ref{dimout_eq}.

\begin{equation}
    H_{out} = \frac{H_{in} - Z+2 \cdot P}{S}+1
\label{dimout_eq}
\end{equation}
Usually, a conv. layer is followed by a \textbf{pooling layer} that aims at reducing the dimensions of the output tensor by locally reducing it with some statistics. A classical approach is to apply a \textit{Max pooling} or an \textit{Average pooling} with a $2\times2$ kernel (we note $Z_{pool}=2$) over the output tensor $Y$ of size $H_{out}\times H_{out}\times C_{out}$ so that the resulting tensor is half the size $(H_{out}/2)\times (H_{out}/2)\times C_{out}$. 
\\

\noindent\textbf{Activation functions ($\sigma$)} inject non-linearity through layers. Typical functions map the output of a dense or conv. layer into specific space like $[0,+\infty[$ (ReLU), $[-1,+1]$ (\textit{Tanh}) or $[0,1]$ (\textit{Sigmoid}, \textit{Softmax}). A widely used function is the Rectified Linear Unit defined as $ReLU(x)=max(0,x)$. The same activation function is applied to all units of a layer and thus can be functionally considered as an independent layer, as in this work. 

\subsection{Model deployment on Cortex-M platforms}

\noindent Several tools are available to deploy DNN on Cortex-M platforms such as TF-LM\footnote{\url{https://www.tensorflow.org/lite/microcontrollers}}, Cube.MX.AI\footnote{\url{https://www.st.com/en/embedded-software/x-cube-ai.html}} (STMicroelectronics), NNoM~\cite{jianjia_ma_2020_4158710} or MCUNet~\cite{lin2020mcunet}. Most of them are based on the CMSIS-NN library from ARM~\cite{lai2018cmsisnn}. In this work, we study implementations based on NNoM with CMSIS-NN as back-end.  

\textbf{NNoM}~\cite{jianjia_ma_2020_4158710} (standing for \textit{Neural Network on Microcontroller}) is an open-source, high-level neural network inference library dedicated to microcontrollers. It allows to easily implement DNN models previously trained on Keras-Tensorflow, while supporting complex structures (e.g. Inception or ResNet). Converted NNoM models are layer-wise quantized to reduce memory footprint. Scaling factor of the quantization scheme is restricted to powers of two, allowing to perform efficient shifting operations rather than divisions. 

Back-end operations can be performed by NNoM's eigenfunctions or by efficient CMSIS-NN ones when compatible.

\textbf{CMSIS-NN}~\cite{lai2018cmsisnn} is a collection of optimized neural network basic operations, developed for Cortex-M processor cores. They enhance performance and reduce memory footprint with different optimisation techniques that depend on the target platform. They handle quantized variables on 8 or 16 bits. Implementations considered hereafter manipulate 8-bit variables. 
Using such variables enables to leverage on Single Instruction Multiple Data (SIMD).

\subsection{Model Extraction}
\noindent The goal of \textit{model extraction} attacks is to steal a \textit{victim model} $M_{W}$ with different possible adversarial goals~\cite{jagielski2020high}. A \textit{task-performance} objective~\cite{orekondy2019knockoff,tramer2016stealing} is to steal the performance of $M_{W}$ to reach equal or better one at lower cost (e.g., save prohibitive training time). In that case, knowledge of (exact) victim model architecture or parameter values is not necessary. 
In a \textit{fidelity} scenario, attacker wants to craft a \textit{substitute model} $M'_{\Theta}$ that mimics $M_W$ as close as possible. $M'$ should provide the same predictions as $M$ (correct and incorrect ones). This 
\textit{similarity} between $M$ and $M'$ is typically defined by measuring the agreement at the label-level~\cite{jagielski2020high}, i.e. $\argmax(M'_\Theta(x)) = \argmax(M_W(x))$ for every $x$ sampled from a target distribution $\mathcal{X}$. A more complex and optimal objective (\textit{Functionally Equivalent Extraction}) targets equal predictions ($M'(x)=M(x), \forall x\in \mathcal{X}$). Importantly, the strongest \textit{Exact Extraction} attack ($\mathcal{A}_M = \mathcal{A}_{M'}$, $W=\Theta$) is impracticable by simply exploiting input/output pairs from victim model \cite{jagielski2020high}. \textit{Fidelity}-oriented scenarios receive a growing interest because of challenging extraction processes, more essentially for the parameters. When dealing with parameter extraction, a usual assumption is that the adversary knows $\mathcal{A}_M$.
 
This is the case for cryptanalysis-like approaches~\cite{carlini2020cryptanalytic,jagielski2020high}, active learning techniques~\cite{papernot2017practical,tramer2016stealing} and recent efforts relying on physical attacks such as side-channel (SCA)~\cite{batina2019csi,joud2023practical} or fault injection (FIA)~\cite{rakin2022deepsteal,hector2023fault} analysis. 
Interestingly, whatever the adversarial goal, victim model architecture is a crucial information: it is compulsory for fidelity scenarios, and its knowledge significantly strengthen attacker's abilities to succeed in task-performance ones~\cite{orekondy2019knockoff}.

\section{Related works and contributions}
\label{soa}

\begin{table}[t]
\caption{Related State-of-the-Art works. n.s.: Not Specified}
\label{tab:archi_xtr_soa}
\centering
\renewcommand{\arraystretch}{1.2}
\begin{tabular}{cccr}
\toprule
Attack &  Physical target &  Targeted models & Used technique \\ 
\midrule
\cite{duddu2018stealing} & MLaaS & CNN & TA \& Regression \\
\cite{yu2020deepem,yli2021extraction} &  FPGA &  BNN &  SEMA \\
\cite{luo2022nnrearch} &  FPGA &  CNN \& ResNet &  SEMA \\
\cite{chmielewski2021reverse} &  GPU &  CNN & SEMA \& TA \\
\cite{batina2019csi} &  µC &  MLP \& CNN & CEMA \\
\cite{xiang2020open} & µC & CNN & SPA \& ML \\
\textit{ours} & µC & MLP \& CNN & SEMA only \\
\bottomrule
\end{tabular}
\end{table}

\noindent A growing number of works investigates architecture extraction with physical attacks. Proposed techniques are essentially based on Side-Channel Analysis (SCA) such as Timing Analysis (TA), Simple Power/EM Analysis (SPA/SEMA) or Cache Attacks (CA) targeting various physical targets (FPGA, microcontrollers (µC), GPU and cloud services hosting DNN models referenced as MLaaS for Machine Learning as a Service). These works are summed up in Table~\ref{tab:archi_xtr_soa}. Usually, architecture and parameters are extracted distinctively, however Batina \etal~propose in~\cite{batina2019csi} to identify layer boundaries while performing parameters extraction with Correlation EM Analysis (CEMA). Correlation scores are used to deduce if currently targeted neuron belongs to the same layer as previous targeted one or not. In other words, if CEMA \textit{fails}, then that means we pass through the next layer. 
Since parameters extraction with CEMA is highly challenging, as detailed in~\cite{joud2023practical}, this method raises practical issues and, to the best of our knowledge, has not been fully demonstrated. Other SCA techniques have been used like TA as in~\cite{duddu2018stealing} to recover model numbers of layers. SCA are also occasionally combined with other approaches such as learning-based clone reconstruction~\cite{yu2020deepem,rakin2022deepsteal} or ML-based classification of traces among a limited set of architectures~\cite{xiang2020open}\footnote{In~\cite{xiang2020open}, authors consider 4 variants of AlexNet, Inceptionv3, ResNet-50 and 101, for a total of 16 architectures.}. More recently,~\cite{luo2022nnrearch} presents a complete analysis of the impact of FPGA implementations of CNN on the related electromagnetic activity and proposes and evaluates an obfuscation-based countermeasure.

In this work, our goal is to highlight leakages related to the architecture of embedded DNN models. We are considering models implemented on a 32-bit microcontroller thanks to NNoM library~\cite{jianjia_ma_2020_4158710} that relies on the widely used CMSIS-NN module~\cite{lai2018cmsisnn}. Both of these tools are open-source and as such, we logically consider the attacker has a total access to their code. 
Acquisitions presented all along this work are obtained with 16 averaged traces all acquired while performing the inference algorithm from a single input belonging to the test set (not used during model training). We deliberately set in such a minimalist setting to assess the level of information that could be extracted by an adversary that has almost no prior knowledge about the victim model and limited experimental data available. This work is not intended to be exhaustive with regards to the extraction of every hyper-parameter of every possible layer types. As most of reference papers in model extraction field, we focus on most common layers and related parameters, i.e. by targeting MLP and CNN models. Furthermore, all our traces, implementations (and complementary results) are publicly available in order to foster new experiments on this topic that we claim to be highly critical in the actual large-scale model deployment context.

\section{Experimental setup}
\label{exp_setup}
\subsection{Models and datasets} 
\begin{figure}[t!]
\includegraphics[width=\textwidth]{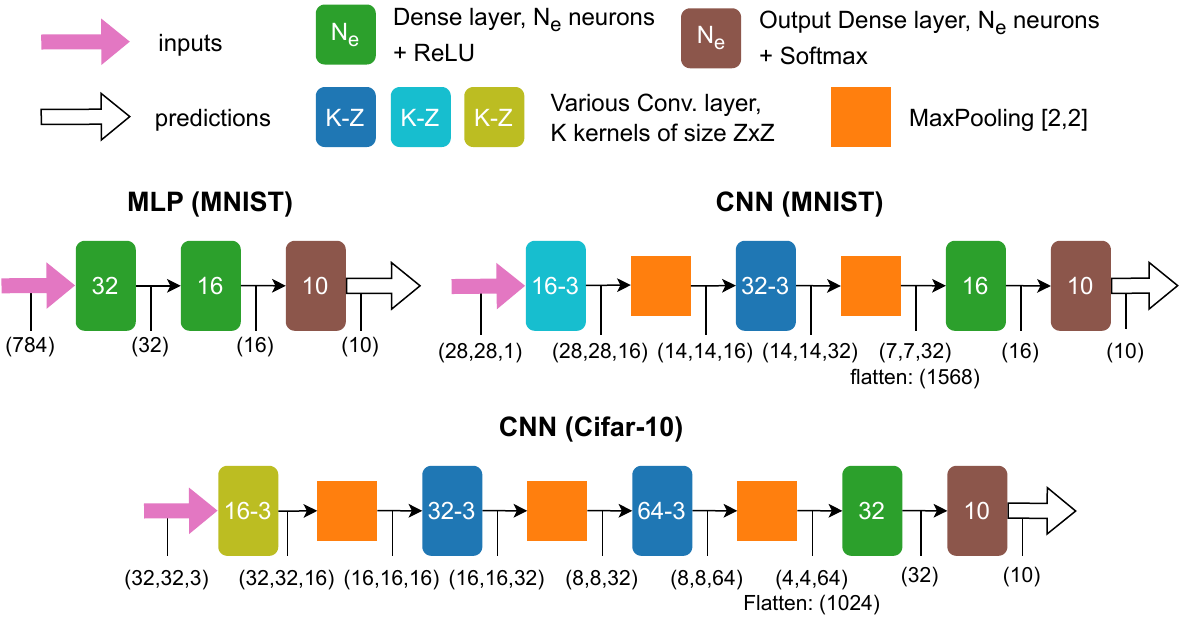}
\caption{Detailed architecture of considered models. Various conv. implementations are detailed in Section~\ref{exp_section:conv_overview}}
\label{diagram_models_archi}
\end{figure}
\noindent We use two traditional benchmarks MNIST\footnote{\url{http://yann.lecun.com/exdb/MNIST/}} (28x28 digit grayscale images, 10 labels) and Cifar-10\footnote{\url{https://www.cs.toronto.edu/~kriz/cifar.html}} (32x32 color images, 10 labels). Therefore, we have $H_{in} = 28$ and $C_{in}=1$ for MNIST and $H_{in} = 32$ and $C_{in}=3$ for Cifar-10. For MNIST, we trained a MLP and a CNN with the TensorFlow platform that achieved respectively $96\%$ and $98\%$ on the test set. For Cifar-10, we trained a CNN that reached $76\%$ on the test set. Architectures of these models are illustrated in Fig.~\ref{diagram_models_archi}. Colors are used to identify the layer and the shape of each output tensor is shown under each layer. Dense layers (\textcolor{Green}{green}) have $N_e$ neurons and convolutional layers (\textcolor{Blue}{blue}, \textcolor{olive}{olive} and \textcolor{cyan}{cyan}) have $K$ square kernels of size $Z$. Classical stride ($=1$) and padding value (\textit{same}) are used, they don't impact the output shape. We only use Max pooling layer (\textcolor{orange}{orange}) with a kernel size of 2 so that the output tensor is reduced by half. Used activation function is ReLU, except for the last layer with SoftMax function.   

\subsection{Target device and setup}
\noindent Our experimental platform is an ARM Cortex-M7 based STM32H7 board that can embed state-of-the-art models thanks to its large memories (2 MBytes of Flash and 1 MByte of SRAM).  
For this first work, interruptions are disabled during model inferences as well as cache optimization available on the board. We measured EM emanations coming from the \textbf{unopened} chip with a Langer probe (EMV-Technik LF-U 2,5, freq. range from 100 kHz to 50 MHz) that is connected to an amplifier (Fento HVA-200M-40-F) with a 200 MHz bandwidth and 40 dB gain. Acquisitions are collected thanks to a Lecroy WavePro 604HD-MS oscilloscope. Additional specifications are in the public repository.

\section{Threat model}
\label{threat_model}
The attacker aims to recover the architecture of an unknown quantized victim model ($M_W$) with as much detail as possible. Table~\ref{tab:info_to_extract} lists all the information the adversary wants to extract thanks to EM traces.
The attack context corresponds to a particular black-box setting. Indeed, the adversary has no knowledge of model architecture nor parameters but is aware of the task performed by the model and the usage of CMSIS-NN module. Adversary is also expected to have appropriate Deep Learning expertise including the most classical and logical layer sequences. The attacker is able to acquire EM side-channel information leaking from the board embedding the targeted DNN model. However, we consider the attacker does not perturb program execution and collect only traces stemming from usual inferences. As we assume a minimal practical setting, adversary is restricted to a single test input. Exploited trace results from the averaging of few inferences with this specific input\footnote{Raw traces are available in the public repository.}.

\begin{table}[h!]
\caption{Adversarial objective: list of the hyper-parameters to extract}
\label{tab:info_to_extract}
\centering
\begin{tabular}{llccllc}
\toprule
Target & Parameters & Notation &\text{ } & Target & Parameters & Notation \\
\midrule

\multirow{2}{4em}{$\mathcal{A}_M$}
 & \# layers      & $L$ & & \multirow{4}{6em}{Conv. Layer}       & Output shape    & $H_{out}$ \\
                & Type of layers & $\varnothing$ & &                 & \# kernels      & $K$ \\
                \cmidrule{1-3}
Dense Layer     & \# neurons     & $N_e$ &&                  & Kernel size     & $Z$ \\ \cmidrule{1-3}
\multirow{2}{4em}{MaxPool}         & Output shape   & $H_{out}$   &&                  & Stride, Padding & $S,P$ \\
\cmidrule{4-7}
                & Filter size    & $Z_{pool}$    & &Activation Layer & ReLU or not     & $\varnothing$ \\

\bottomrule
\end{tabular}%
\end{table}

Furthermore, we want to point out that in such context, the adversary has access to every needed resources for profiling attacks. Target board is known as well as CMSIS-NN back-end usage, allowing to set a dictionary linking layers with various characteristics to their corresponding EM activity. Simple pattern detection tools could also be used to speedup extraction process. However, usage of such tools or profiling techniques is not in the scope of this work.

\section{Layers analysis}
\label{exp_section}
At the model scale, inference is a feedforward process with the computation of each layer performed one after the other. The output tensor of one layer becomes the input tensor of the next one. In this section, we focus at layer scale with a two-step methodology. 
First, we analyze the CMSIS-NN implementation of each layer of our models to reveal repetitions of regular computation blocks that should appear in our EM traces. More particularly, we aim to link these \textit{countable} patterns to hyper-parameters of the layer. Second, we experimentally evaluate if these theoretical assumptions are confirmed in our EM traces and assess the complexity of the extraction and potential limitations.

Our approach is based on the principle that the attack is performed according to the computational flow: when the adversary targets layer $l$, we suppose that analysis of layer $l-1$ is complete and thus its output tensor shape is known, meaning that input tensor shape of layer $l$ is mastered as well\footnote{Obviously, we expect that the size of the inputs feeding the model is known by the adversary (e.g., $28\times28$ for MNIST and $32\times32\times3$ for Cifar-10).}.

\subsection{Convolutional layer}
\label{exp_section:conv}

\noindent CMSIS-NN convolution functions are all based on the same two-step process, as presented in Alg.~\ref{alg_conv_output_size}: Im2col and General Matrix Multiply (GeMM) algorithms. Im2col is a standard optimization trick that implements a convolution through a matrix product rather than several dot-products as defined in Eq.~\ref{eq_cnn}. The trick is to prepare all the local areas obtained from sliding window into column vectors and expand the kernel values into rows. Then, convolution is equivalent to a single matrix multiplication that allows an important execution speedup but at the expense of an increased memory footprint. To reduce the latter, CMSIS-NN iteratively performs Im2col with small sets of column vectors~\cite{lai2018cmsisnn}. Depending on input and output tensors size, three different functions are proposed\footnote{Equivalent functions are available for non-squared input tensor.}:

\begin{itemize}
    \item \texttt{arm\_convolve\_HWC\_q7\_fast()} (\textcolor{blue}{blue conv. layers} in Fig.~\ref{diagram_models_archi}) is for $C_{in}$ multiple of 4 (due to SIMD read and swap) and $C_{out}$ multiple of 2 (due to matrix multiplication applied on $2\times 2$ elements). The computation is speedup for the padding management by splitting the input tensor into $3\times3$ patches. 
    \item \texttt{arm\_convolve\_HWC\_q7\_RGB()} (\textcolor{olive}{olive conv. layers} in Fig.~\ref{diagram_models_archi}) is exclusively optimized for 3 channels inputs (hard-coded condition checks). 
    \item \texttt{arm\_convolve\_HWC\_q7\_basic()} (\textcolor{cyan}{cyan conv. layers} in Fig.~\ref{diagram_models_archi}) is used otherwise and has a very similar structure as the RGB variant. 
\end{itemize}

\label{exp_section:conv_overview}
\subsubsection{Output shape ($H_{out}$)}

\begin{algorithm}[h!]\scriptsize
\caption{General conv. implementation}\label{alg_conv_output_size}
\begin{algorithmic}[1]

\Require $I_{in}$ Input tensor of size $H_{in}^2 \cdot C_{in}$, $ker$ (Kernel tensor), $S$, $P$, $I_{out}$ Output tensor of size $H_{out}^2 \cdot C_{out}$
\Ensure Filled $I_{out}$

\For{$i_y \gets 0,\ i_y < H_{out},\ i_y++1$} \Comment{\textcolor{ForestGreen}{Iterate over $H_{out}$ ($y$-axis)}}
\For{$i_x \gets 0,\ i_x < H_{out},\ i_x++1$} \Comment{\textcolor{ForestGreen}{idem ($x$-axis)}}

\State buff$_{in} \gets im2col(I_{in}, i_y, i_x, S, P, H_{in}, C_{in})$ \Comment{\textcolor{ForestGreen}{Apply im2col conversion}}

\If{$len($buff$_{in}) == 2 \times C_{in} \times Z^2$} \Comment{\textcolor{ForestGreen}{Check if 2 input columns are set}}
\State $GeMM($buff$_{in}, ker, C_{out}, C_{in} \times Z^2, I_{out})$ \Comment{\textcolor{ForestGreen}{Perform matrix-multiplication}}
\State buff$_{in} \gets 0$ \Comment{\textcolor{ForestGreen}{Buffer reset}}
\EndIf

\EndFor
\EndFor
\end{algorithmic}
\end{algorithm}

\paragraph{Code analysis.} Whatever the (optimization) differences between the three  implementations listed above, extraction of $H_{out}$ relies on the same principle. As presented in Alg.~\ref{alg_conv_output_size}, outer loops iterate over the tensor size $H_{out}$ (reminder: we consider square inputs) to run the core computations with the im2col trick then matrix multiplication (GeMM). These computations are time consuming and are likely to induce clear visible EM activities, especially GeMM step (Alg.\ref{alg_conv_nb_kernel} described after). GeMM is called every two iterations (line 4, \texttt{if} statement) when the buffer buff$_{in}$ is filled with two \textit{input columns} thanks to im2col (line 3). As a result, GeMM function (line 5) is called $H_{out} \times H_{out}/2$ times during Alg.~\ref{alg_conv_output_size} execution. If we set $Np$ as the number of regular patterns resulting from GeMM function over the part of the EM trace corresponding to the targeted conv. layer, then we can link $Np$ to $H_{out}$ as: $N_{p} = H_{out} \times H_{out}/2$, i.e. $H_{out} = \sqrt{2 \times N_{p}}$.

\begin{figure}[h]
\centering
\begin{subfigure}[t]{0.49\textwidth}
    \includegraphics[width=1\textwidth]{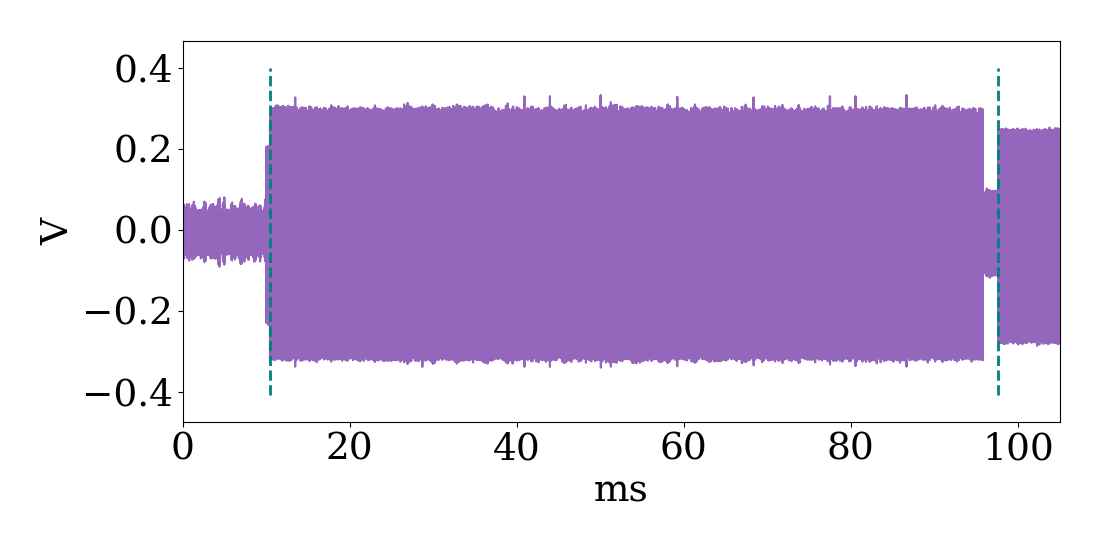}
    \caption{$1^{st}$ conv. (\texttt{basic}), MNIST CNN, 392 expected patterns}
    \label{fig_conv0_MNIST}
\end{subfigure}\hspace{\fill}
\begin{subfigure}[t]{0.49\textwidth}
    \includegraphics[width=1\textwidth]{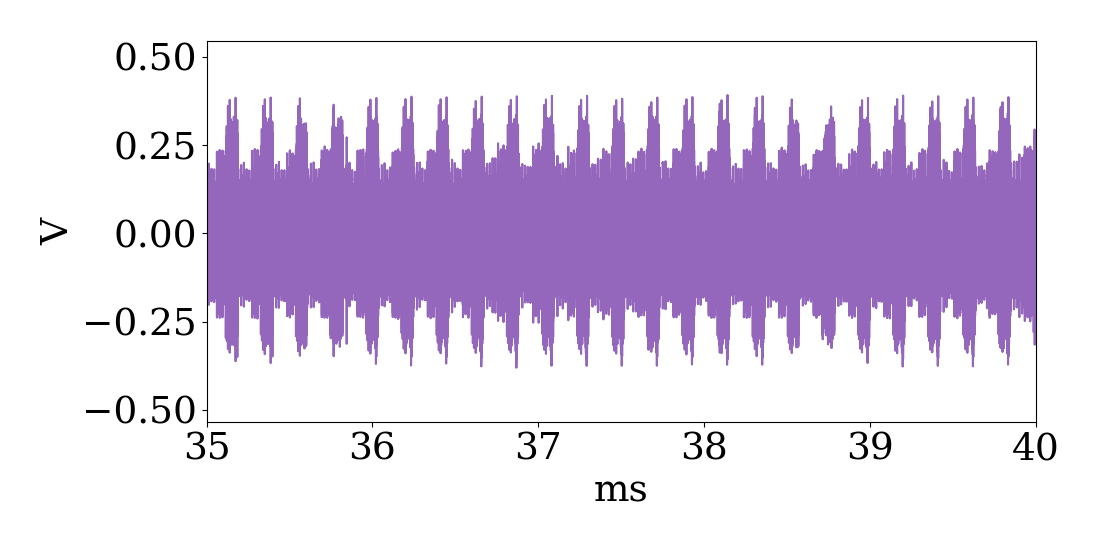}
    \caption{Zoom in on (a) 23 visible patterns}
    \label{fig_conv0_MNIST_zoom}
\end{subfigure}

\vspace{10pt}

\begin{subfigure}[t]{0.49\textwidth}
    \includegraphics[width=1\textwidth]{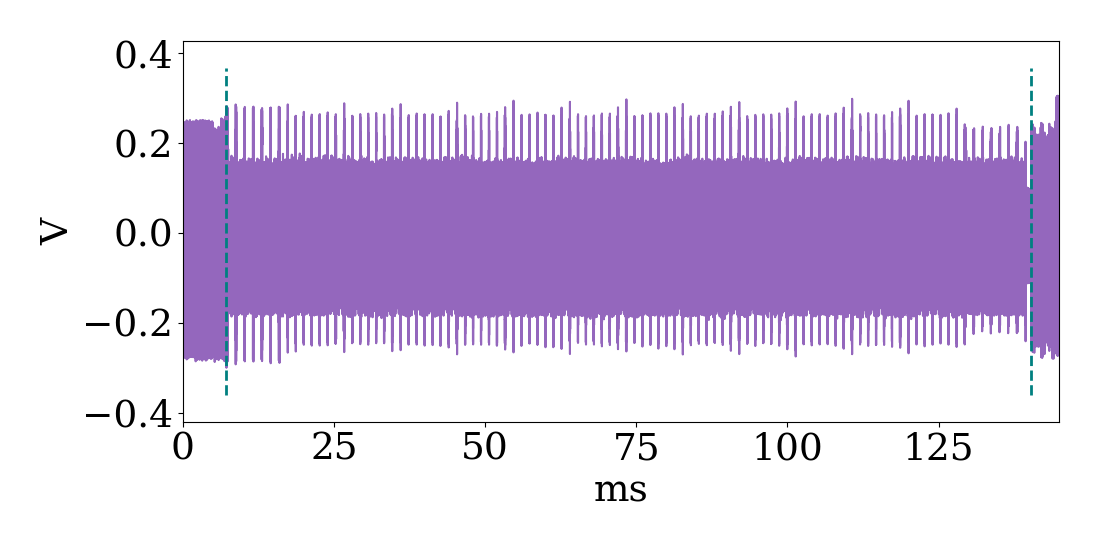}
    \caption{$2^{nd}$ conv. (\texttt{fast}), MNIST CNN, 98 visible patterns}
    \label{fig_conv1_MNIST}
\end{subfigure}\hspace{\fill}
\begin{subfigure}[t]{0.49\textwidth}
    \centering
    \includegraphics[width=1\textwidth]{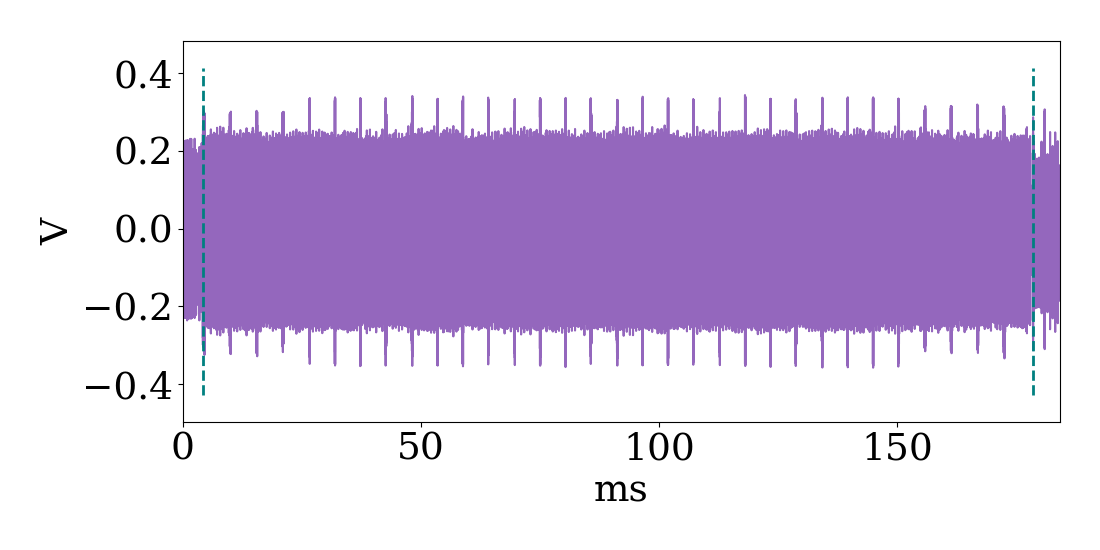}
    \caption{$3^{rd}$ conv. (\texttt{fast}), Cifar-10 CNN, 32 visible patterns\vspace{5pt}}
    \label{fig_conv2_CIFAR}
\end{subfigure}

\caption{Overviews of conv. layers EM activity related to $H_{out}$}
\label{trace_conv_overview}
\end{figure}

\paragraph{Observations and limitations.} Thanks to this analysis, we assess our traces for each conv. layer of the two CNN models (MNIST and Cifar-10) and observe the correct number of patterns $Np$ as noticed in Tab.~\ref{tab:Conv_out_dim}.
Fig~\ref{fig_conv0_MNIST} illustrates the first conv. layer for MNIST with $Np=392$ patterns (Fig.~\ref{fig_conv0_MNIST_zoom} gives a zoom on 23 patterns between 35 and 40 ms). Fig~\ref{fig_conv1_MNIST} and~\ref{fig_conv2_CIFAR} illustrate \texttt{fast} conv. layers of either MNIST or Cifar-10 CNN with respectively $Np=98$ and 32 patterns. After counting $Np$, we easily deduce $H_{out}$ values.
According to observed layers, number of patterns can be important and hard to count \textit{by hand}. However, since these patterns are clear and accurate (see Fig.~\ref{fig_conv0_MNIST_zoom} and \ref{fig_conv2_CIFAR}), an attacker could take advantage of basic pattern detection tools (out of the scope of this work) to make the counting easier. 

Next, we zoom in one pattern to focus on the GeMM function and look for other regular patterns that may be related to other hyper-parameters, more particularly the number of kernels $K$ and their size $Z$. 
\subsubsection{Number of kernels ($K$)} 
\paragraph{Code analysis.} Matrix-multiplication computation is described in Alg.~\ref{alg_conv_nb_kernel}. It mainly consists of an outer \texttt{for} loop iterating over the number of output channels ($C_{out}$) divided by 2, referenced as \texttt{rowCnt} and defined at line 1. We remind that for a conv. layer: $C_{out} = K$. Thus, being able to count the iterations of the \texttt{for} loop  (lines 2-9) enables to recover $K$. If $N_p$ is the number of regular EM patterns resulting from computations inside this loop, then we assume that $K = 2 \times N_p$.

\noindent\begin{minipage}[h]{\linewidth}

\begin{minipage}[h]{0.64\linewidth}

\begin{algorithm}[H]\scriptsize
\caption{Matrix-product for conv. (\textbf{GeMM})}
\label{alg_conv_nb_kernel}
\begin{algorithmic}[1]

\Require buff$_{in}$, $ker$, $C_{out}$, $C_{in} \times Z^2$, $I_{out}$, $bias$ (bias tensor)
\Ensure Partly filled $I_{out}$

\State $rowCnt \gets C_{out} >> 1$ \Comment{\textcolor{ForestGreen}{Set $rowCnt = K/2$}}

\For{$rowCnt > 0,\ rowCnt--1$} \Comment{\textcolor{ForestGreen}{Iterate over $K/2$}}
\State $sum, sum1, sum2, sum3 = init\_sum(bias, C_{in} \times Z^2)$
\State $colCnt \gets (C_{in} \times Z^2) >> 2$ \Comment{\textcolor{ForestGreen}{Set $colCnt$ as in Eq.~\ref{eq_ker_size}}}

\For{$colCnt>0,\ colCnt--1$} 
\State $simd\_mac(sum, sum1, sum2, sum3, $buff$_{in}, ker)$
\EndFor

\State $apply\_mac(sum, sum1, sum2, sum3, C_{out}, I_{out})$
\EndFor

\State $Manage\_remainder\_if\_any(C_{out}, $buff$_{in}, bias, C_{in} \times Z^2)$

\end{algorithmic}
\end{algorithm}

\end{minipage}
\hfill
\begin{minipage}[h]{0.3\linewidth}
    \captionof{table}{$H_{out}$ and expected \# patterns}
    \begin{tabular}{c|cccc}
    \toprule
    \multicolumn{1}{l|}{} & \multicolumn{2}{c}{Mnist}     & \multicolumn{2}{c}{Cifar-10}  \\
    Layer                & $H_{out}$     & $N_{p}$ & $H_{out}$     & $N_{p}$ \\
    \midrule
    1                    & 28            & 392           & 32            & 512           \\
    2                    & 14            & 98            & 16            & 128           \\
    3                    & $\varnothing$ & $\varnothing$ & 8             & 32\\
    \bottomrule
    \end{tabular}
    \label{tab:Conv_out_dim}
\end{minipage}
\end{minipage}

\begin{figure}[h]
\centering
\begin{subfigure}[t]{0.49\textwidth}
    \includegraphics[width=1\textwidth]{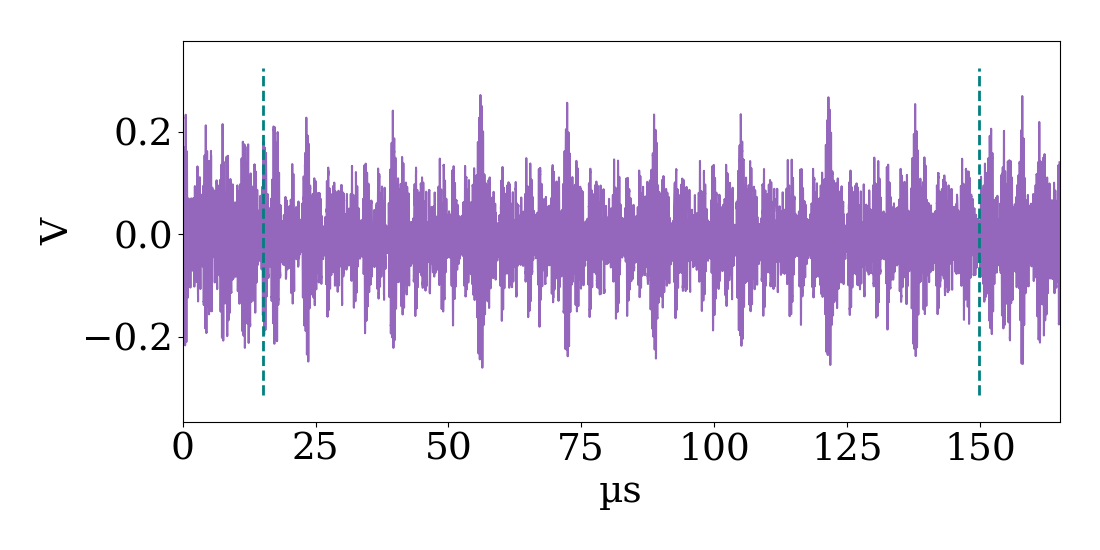}
    \caption{$1^{st}$ conv., 8 visible patterns, $K=16$\vspace{10pt}}
    \label{K_MnistConv1}
\end{subfigure}\hspace{\fill}
\begin{subfigure}[t]{0.49\textwidth}
    \includegraphics[width=1\textwidth]{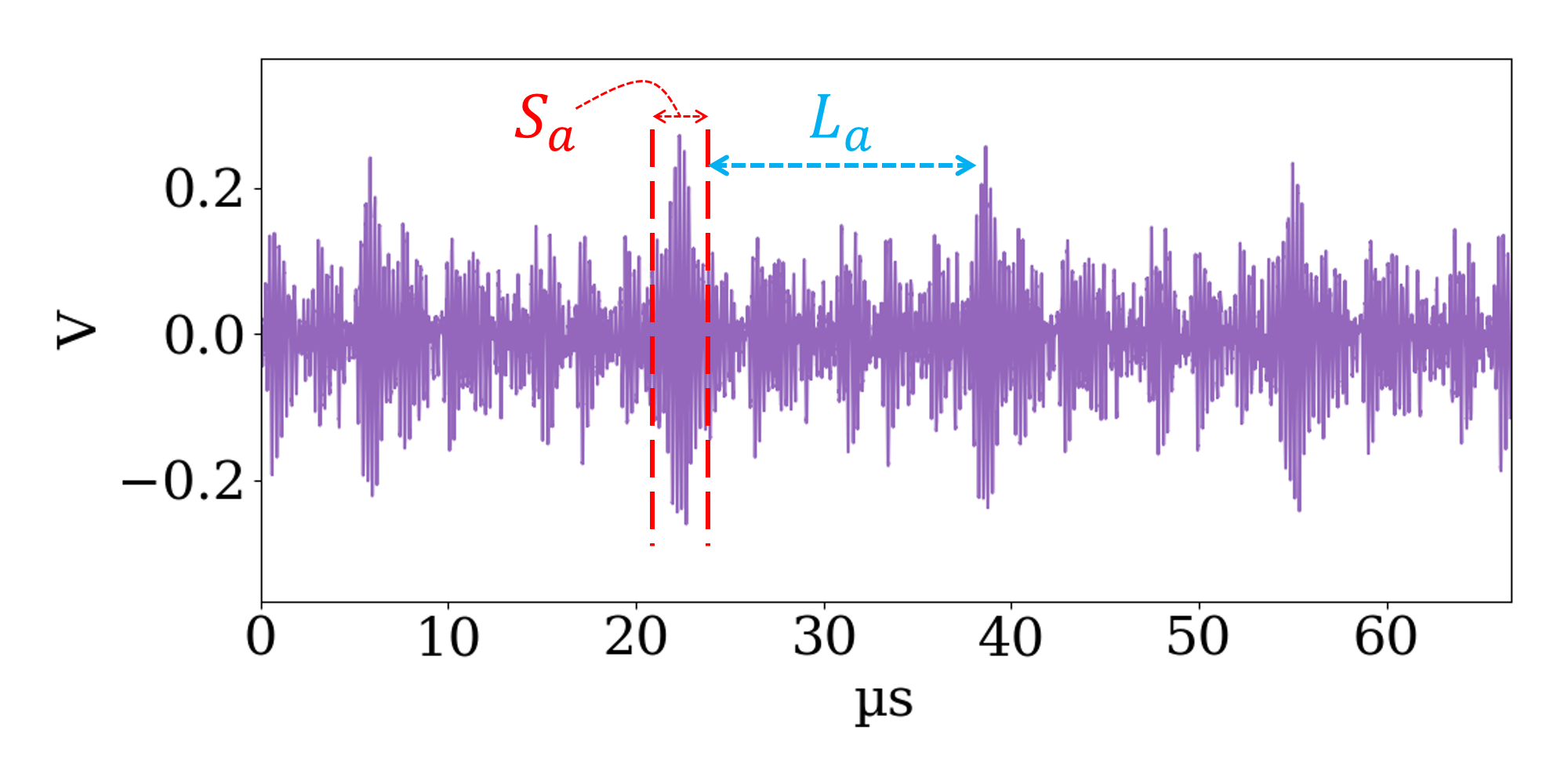}
    \caption{Zoom over $1^{st}$ conv.}
    \label{zoomedK_MnistConv1}
\end{subfigure}

\begin{subfigure}[t]{0.49\textwidth}
    \includegraphics[width=1\textwidth]{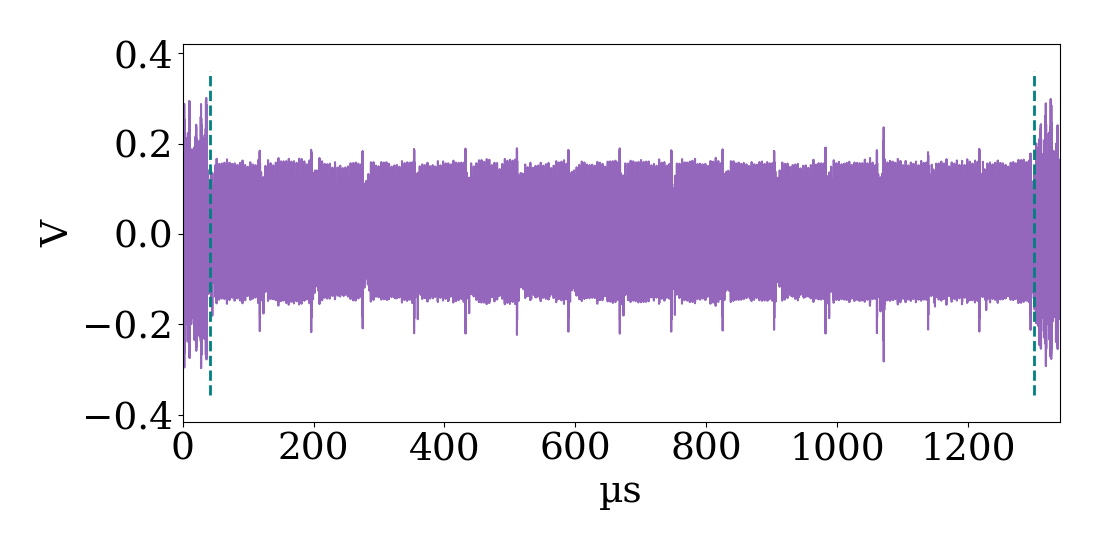}
    \caption{$2^{nd}$ conv., 16 visible patterns, $K=32$\vspace{10pt}}
    \label{K_MnistConv2}
\end{subfigure}\hspace{\fill}
\begin{subfigure}[t]{0.49\textwidth}
    \includegraphics[width=1\textwidth]{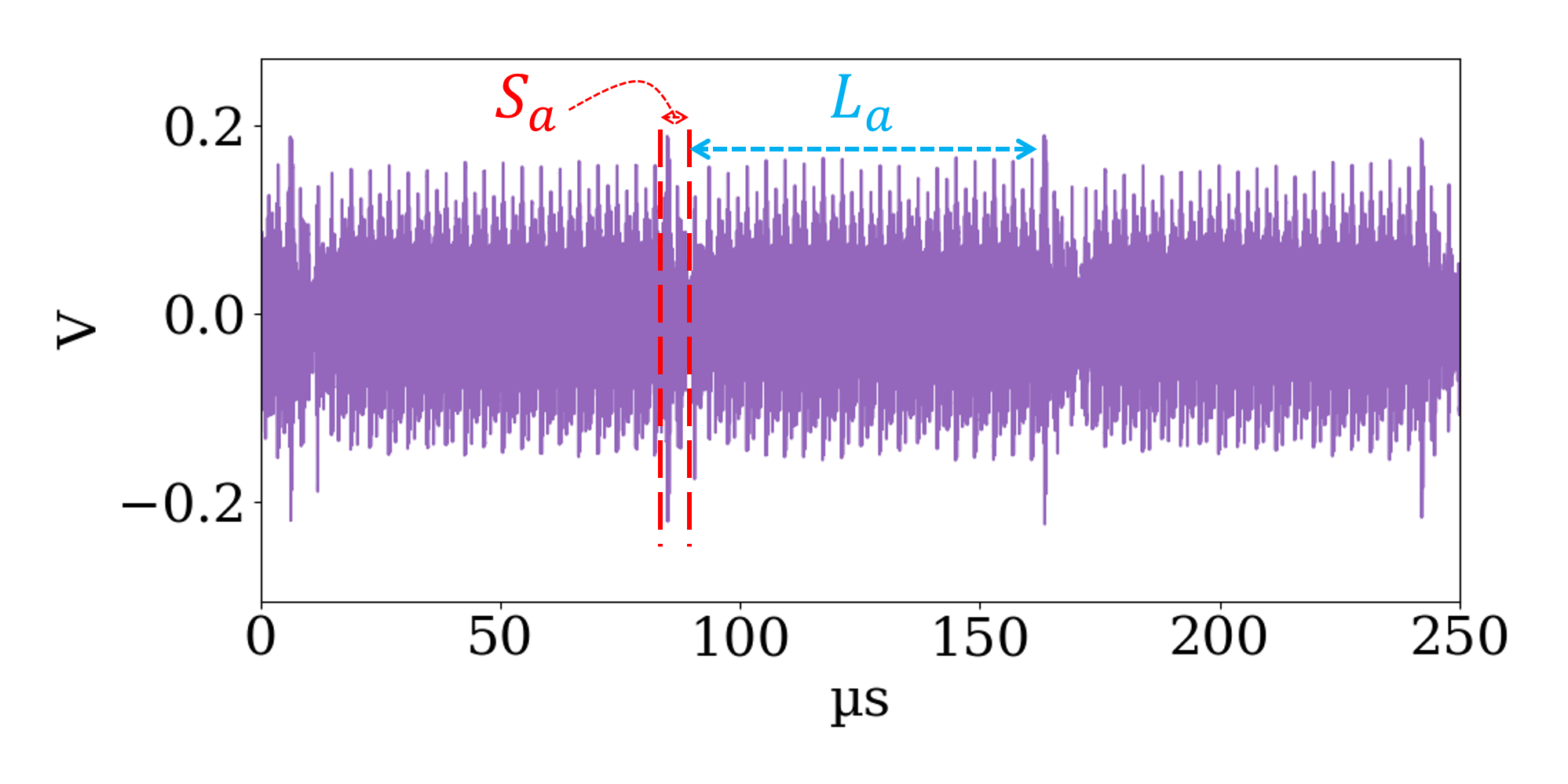}
    \caption{Zoom over $2^{nd}$ conv.\vspace{5pt}}
    \label{zoomedK_MnistConv2}
\end{subfigure}

\caption{Single GeMM execution traces with zoom for MNIST conv. layers}
\label{trace_number_kernel}
\end{figure}

\paragraph{Observations.} We evaluate the number of regular patterns appearing on our traces when zooming in the first pattern we previously extracted $H_{out}$ from. 
For each conv. layer, we observe groups of new regular patterns composed of the expected number $N_p$. Furthermore, these groups are observed throughout the EM activity related to the entire conv. layer, corresponding to every call of the GeMM function of Alg.~\ref{alg_conv_output_size}. These regular patterns are composed of a spike $(Sa)$ followed by a segment of lower frequency activity $(La)$ (Fig.~\ref{zoomedK_MnistConv1} and~\ref{zoomedK_MnistConv2}). Fig~\ref{K_MnistConv1} and~\ref{K_MnistConv2} represent targeted outer \texttt{for} loop iterations for the first two conv. layers of MNIST CNN with respectively $N_p = 8$ and 16, that corresponds to $K=16$ and 32 kernels. Such observations have been successfully made for every other conv. layers for both Cifar-10 and MNIST CNN models.

\subsubsection{Size of kernel ($Z$)}
\paragraph{Code analysis.} Like the number $K$ of kernels, their size $Z$ is also manipulated in Alg.~\ref{alg_conv_nb_kernel}. The inner \texttt{for} loop (lines 5-7) iterates over the variable \texttt{colCnt} which directly depends on $Z$. This part represents the core computation of convolution (with SIMD-based accumulations, line 6). \texttt{colCnt} assignation (line 4) is as Eq.~\ref{eq_ker_size}:
\begin{equation}
    colCnt = \tfrac{1}{4}(C_{in}\times Z^2)
\label{eq_ker_size}
\end{equation}
Thus, as before, we expect that if these very regular computations result in regular EM patterns, then we can estimate $Z$. However, such a statement is based on the knowledge of $C_{in}$. As previously mentioned, the attacker is supposed to master the dimensions of the output tensor of the previous layer, meaning that he knows $C_{in}$. So, if EM activity related to the inner \texttt{for} loop can be distinguished from the rest with $N_p$ regular patterns, then $Z$ is estimated as $\sqrt{(4\times N_p)/C_{in}}$.

\begin{figure}[t!]
\centering

\begin{subfigure}[t]{0.49\textwidth}
    \includegraphics[width=1\textwidth]{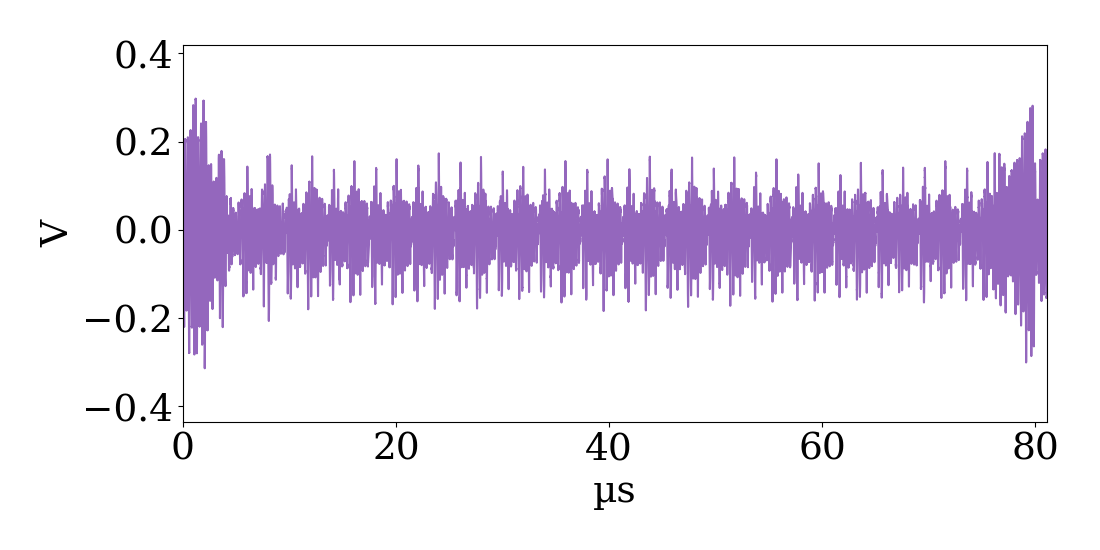}
    \caption{$2^{nd}$ conv., Cifar-10 CNN \\\textit{36 visible patterns with $C_{in}=16$}}
    \label{Z_conv1_CIFAR}
\end{subfigure}\hspace{\fill}
\begin{subfigure}[t]{0.49\textwidth}
    \includegraphics[width=1\textwidth]{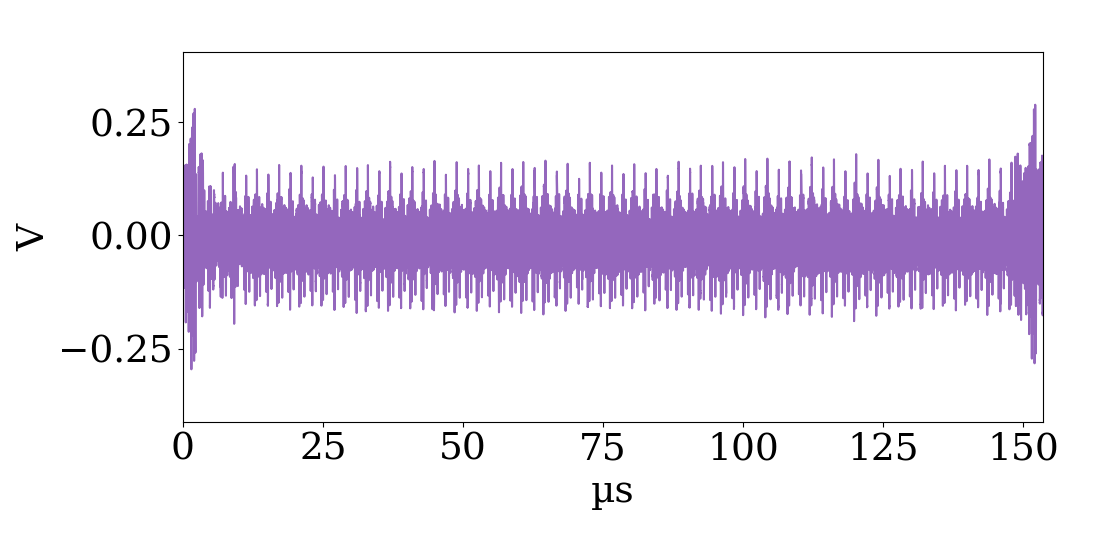}
    \caption{$3^{rd}$ conv., Cifar-10 CNN, \\\textit{72 visible patterns with $C_{in}=32$}\vspace{5pt}}
    \label{Z_conv2_CIFAR}
\end{subfigure}

\caption{Zoom in Matrix-product EM activity with kernels of size $Z=3$}
\label{trace_kernel_size}
\end{figure}

\paragraph{Observations.} When observing EM activity between two spikes defined previously as $La$ (Fig.~\ref{trace_number_kernel}), repetitive patterns can be spotted as in Fig.~\ref{trace_kernel_size}. EM activities related to SIMD-based accumulations of the second and third conv. layer of Cifar-10 CNN are shown respectively in traces~\ref{Z_conv1_CIFAR} and~\ref{Z_conv2_CIFAR}. For the second layer, $C_{in}=16$ and $N_p=36$ patterns can be counted, giving correctly $Z=3$ according to previous assumption. Same correct result is obtained for the third layer with $C_{in}=32$ and $N_p=64$ visible patterns.

\paragraph{Limitations.} It is important to note that dividing $C_{in}\times Z^2$ by 4 can induce remainders to be treated separately afterwards. 
In this case, another EM activity will appear after the \texttt{colCnt} expected patterns. Alternatively, patterns to be seen and counted are becoming pretty small and can be hard to distinguish one another. Moreover, the value of $C_{in}$ strongly impacts the number of patterns to differentiate. As an example, number of input channels in the first convolution layer of MNIST CNN is necessarily equal to 1 as input are grayscale images. With $3\times3$ kernels, only two patterns shall appear, with division remainder treated afterwards that emits different EM activity. In other words, $K$ can be challenging to recover, especially when $C_{in}$ is small.

\subsubsection{Stride ($S$) and Padding ($P$)}
\paragraph{\normalfont{Once}} $H_{in}$, $H_{out}$ and $Z$ are known, based on the fact that $P < Z$ and thanks to Eq.~\ref{dimout_eq}, $S$ and $P$ can be deduced. By computing $S$ for possible $P$ values, only a single integer value will result for $S$.

\subsection{Pooling layer:  Output dimensions ($H_{out}$) and kernel size ($Z_{pool}$)}
\label{exp_section:maxpool}

\paragraph{Code analysis.} CMSIS-NN implementation (Alg.~\ref{alg_maxpool}) of max-pooling relies on two computational blocks. First one (lines 1-6) handles the pooling along $x$-axis with two nested loops iterating over $H_{in}$ and $H_{out}$. From an adversary's point of view, the second block (lines 8-11) is the most interesting since it completes the pooling over the $y$-axis with a single loop over $H_{out}$ only. Pooling operation is classically performed with two steps: (1) selection and allocation of local areas, then (2) statistic computation (here, maximum). Thus, we expect to observe two distinct EM activities related to these separated loops with the last (and shorter) one directly related to $H_{out}$. Then we can deduce $Z_{pool}=H_{out}/H_{in}$.
\begin{algorithm}[h!]\scriptsize
\caption{MaxPool - \texttt{arm\_maxpool\_q7\_HWC}}\label{alg_maxpool}
\begin{algorithmic}[1]

\Require $I_{in}$ input tenor of size $H_{in}^2\cdot C_{in}$, $I_{out}$ output tenor of size $H_{out}^2 \cdot C_{out}$, $P$, $S$, $H_{ker}$
\Ensure Filled $I_{out}$

\For{$i_y \gets 0,\ i_y < H_{in},\ i_y++1$} \Comment{\textcolor{ForestGreen}{Pooling along x axis}}
\For{$i_x \gets 0,\ i_x < H_{out},\ i_x++1$} 
\State $win_{start}, win_{stop} \gets set\_window(i_y, i_x, I_{in}, H_{ker}, P, S)$
\State $compare\_and\_remplace\_if\_larger(win_{start}, win_{stop}, i_y, i_x, I_{in})$
\EndFor
\EndFor
\State $trigger\_up()$ \Comment{\textcolor{ForestGreen}{Pooling along y axis}}
\For{$i_y \gets 0,\ i_y < H_{out},\ i_y++1$} \Comment{\textcolor{ForestGreen}{Directly iterates over $H_{out}$}}
\State $row_{start}, row_{stop} \gets set\_rows(i_y, I_{in}, I_{out}, H_{ker}, P, S)$
\State $compare\_replace\_then\_apply(row_{start}, row_{stop}, I_{in}, I_{out})$
\EndFor
\State $trigger\_down()$
\end{algorithmic}
\end{algorithm}

\begin{figure}[t!]
\centering

\begin{subfigure}[t]{0.49\textwidth}
    \includegraphics[width=1\textwidth]{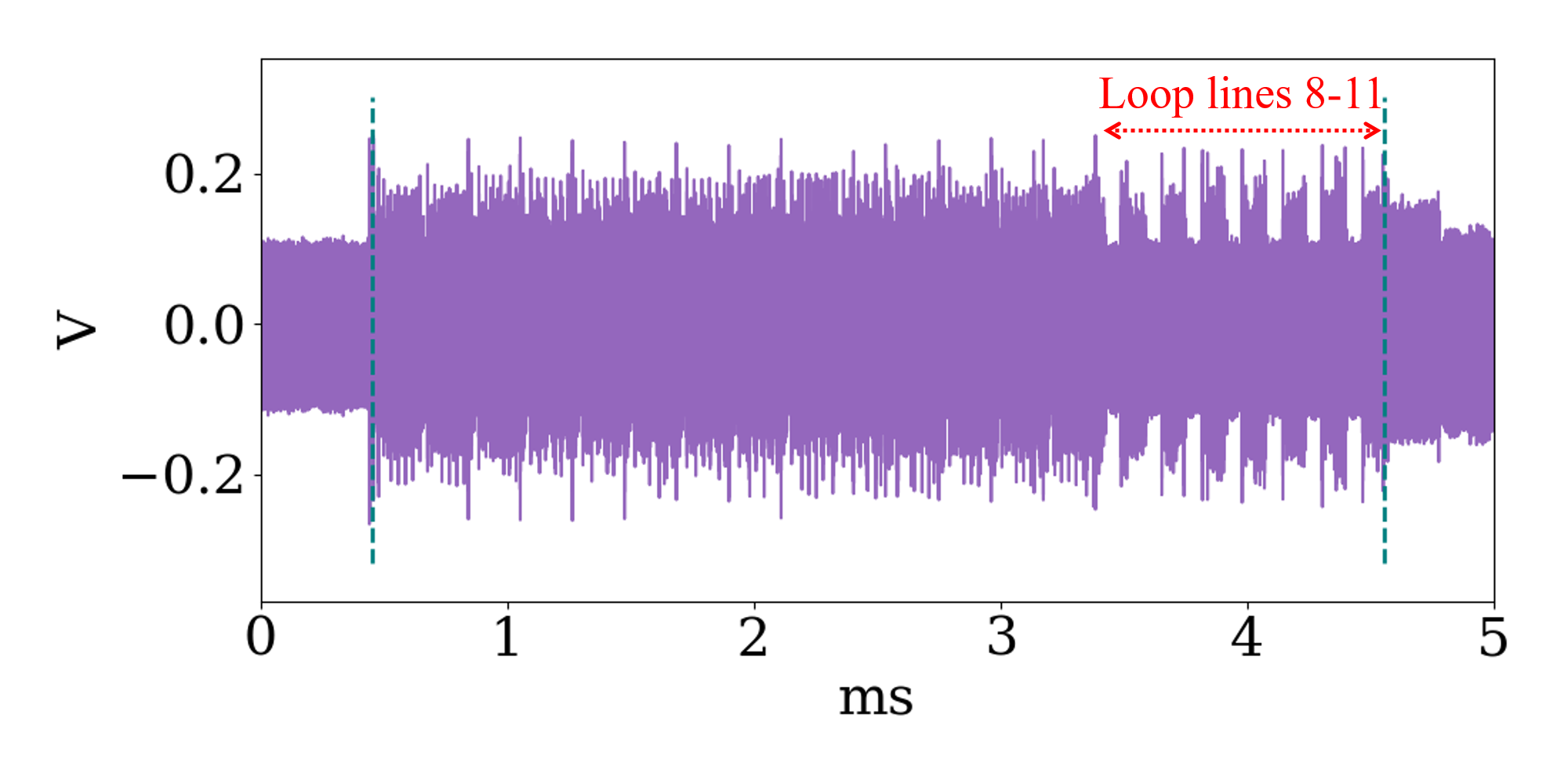}
    \caption{Overview of $2^{nd}$ maxpool of MNIST CNN, \textit{$H_{out}=7$}} 
    \label{mnist_maxpool1_overview}
\end{subfigure}\hspace{\fill}
\begin{subfigure}[t]{0.49\textwidth}
    \includegraphics[width=1\textwidth]{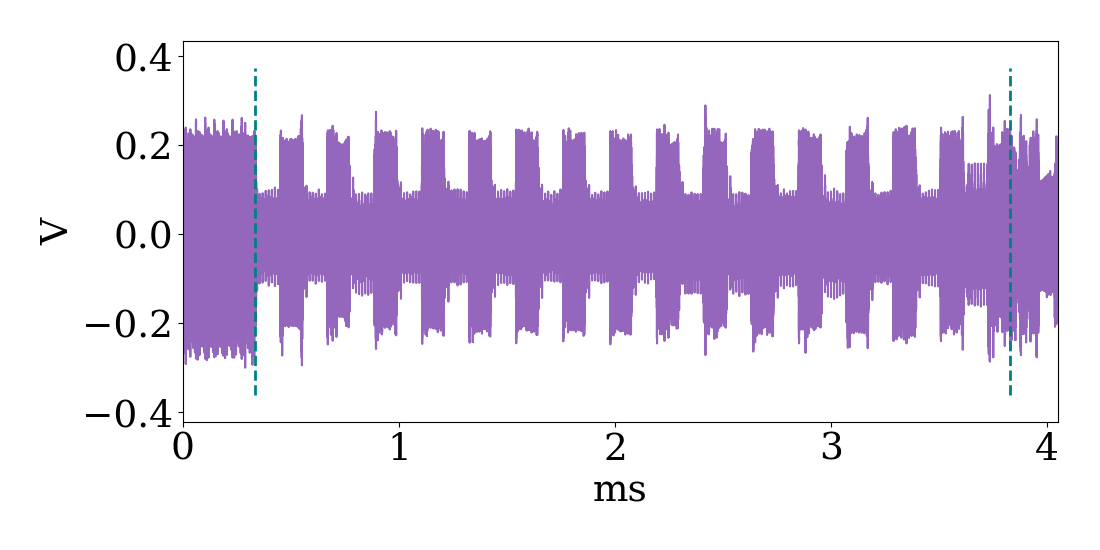}
    \caption{Zoom over $2^{nd}$ loop of $1^{st}$ Maxpool of Cifar-10 CNN, \textit{$H_{out}=16$}, triggers set as in Alg.~\ref{alg_maxpool} at lines 7 and 12\vspace{5pt}}
    \label{cifar10_maxpool0}
\end{subfigure}

%

\caption{MaxPool layers of CNN models}
\label{trace_maxpool_output_vol}
\end{figure}

\paragraph{Observations.}
EM activity of an entire MaxPool layer is represented in Fig~\ref{mnist_maxpool1_overview} ($2^{nd}$ pooling, MNIST CNN). As expected, we observe two distinct blocks of EM emanations, the second one revealing patterns that are more distinguishable. When zooming over this second part, as in Fig~\ref{cifar10_maxpool0} ($1^{st}$ pooling, Cifar-10), these patterns are composed of two segments, one with low amplitude activity followed by another of higher amplitude. We guess that it corresponds to the two steps of the pooling computation. The number $N_p$ of these patterns can easily be hand-counted and matches $H_{out}$ value. We have been able to do equivalent observations for all the MaxPool layers of our CNN models. Experimentally, we note that analyzing the first block to retrieve both $H_{in}$ and $H_{out}$ is feasible but more complex than simply counting $H_{out}$ with the second block. 

\subsection{Dense layer: Number of neurons ($N_e$)}

\paragraph{Code analysis.}
Alg.~\ref{alg_dense} describes function dedicated to dense layers. It is built around a \texttt{for} loop iterating over the \texttt{rowCnt} variable that is the number of neurons $N_e$ divided by 4 (line 1), since neurons are handled by groups of 4. Indeed, \texttt{init\_sum\_with\_bias} function (line 3) sets four sum variables. They are used for weighted sum computation (and bias value addition) through multiply-accumulate SIMD operations (\texttt{\_\_SMLAD}), performed in \texttt{simd\_mac} (line 6). We are likely to observe $N_p$ regular patterns emanating from this loop, with $N_e = 4 \times N_p$. 

\begin{algorithm}[h!]\scriptsize
\caption{Dense layer - \texttt{arm\_fully\_connected\_q7\_opt}}\label{alg_dense}
\begin{algorithmic}[1]

\Require $I_{in}$ input vector of size $H_{in}$, $ker$ weight vector of size $N_e$, $bias$ bias matrix of size $N_e$, $I_{out}$ output vector of size $H_{out}$, $P$, $S$, $H_{ker}$
\Ensure Filled $I_{out}$

\State $rowCnt \gets N_e >> 2$ \Comment{\textcolor{ForestGreen}{Nb. neurons divided by 4}}
\For{$rowCnt > 0,\ rowCnt--1$} \Comment{\textcolor{ForestGreen}{Iterate directly over $N_e/4$}}
\State $sum, sum1, sum2, sum3 = init\_sum\_with\_bias(bias, rowCnt)$
\State $colCnt \gets H_{in} >> 2$
\For{$colCnt > 0,\ colCnt--1$}
\State $simd\_mac(sum, sum1, sum2, sum3, ker, colCnt)$
\EndFor
\State $apply\_mac(sum, sum1, sum2, sum3, rowCnt, I_{out})$
\EndFor

\State $rowCnt \gets N_e\ \&\ 0x3$ \Comment{\textcolor{ForestGreen}{Manage remainders if any}}
\For{$rowCnt > 0,\ rowCnt--1$}
\State $sum = init\_sum\_with\_bias(bias, rowCnt)$
\State $colCnt \gets H_{in} >> 2$
\For{$colCnt > 0,\ colCnt--1$}
\State $mac(sum, ker, colCnt)$
\EndFor
\State $apply\_mac(sum, rowCnt, I_{out})$
\EndFor

\end{algorithmic}
\end{algorithm}

\begin{figure}[t!]
\centering
\begin{subfigure}[t]{0.49\textwidth}
    \includegraphics[width=1\textwidth]{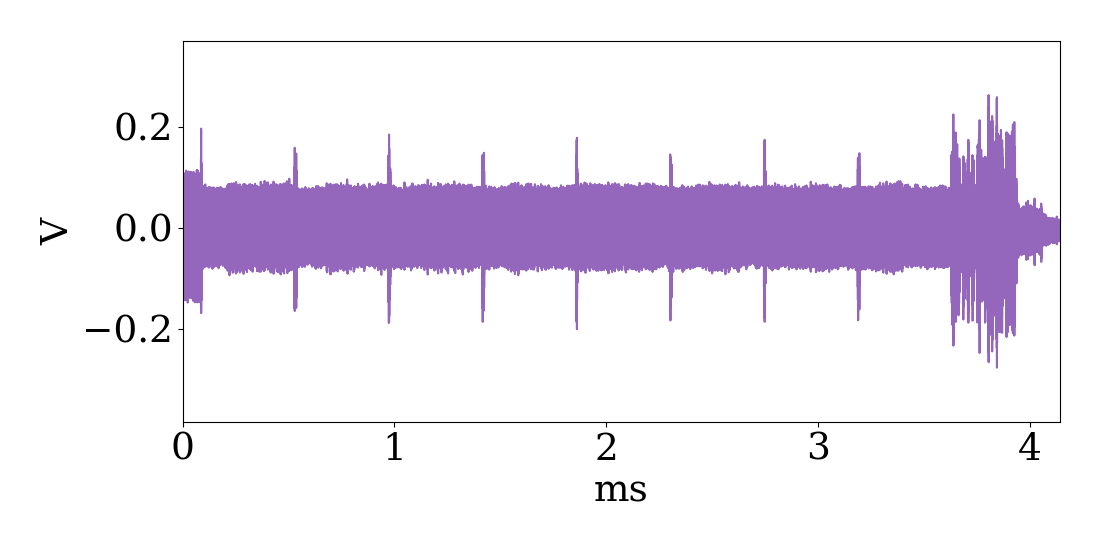}
    \caption{$1^{st}$ dense layer, MNIST MLP, $N_e=32$}
    \label{fig_MNIST-MLP_dense_1}
\end{subfigure}\hspace{\fill}
\begin{subfigure}[t]{0.49\textwidth}
    \includegraphics[width=1\textwidth]{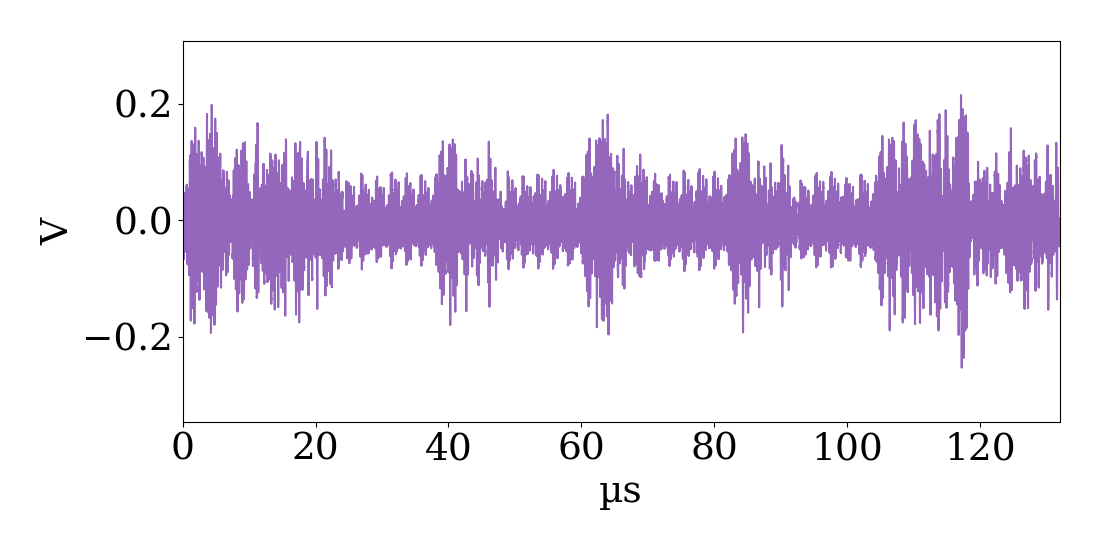}
    \caption{$2^{nd}$ dense layer, MNIST MLP, $N_e=16$\vspace{10pt}}
    \label{fig_MNIST-MLP_dense_2}
\end{subfigure}

\begin{subfigure}[t]{0.49\textwidth}
    \includegraphics[width=1\textwidth]{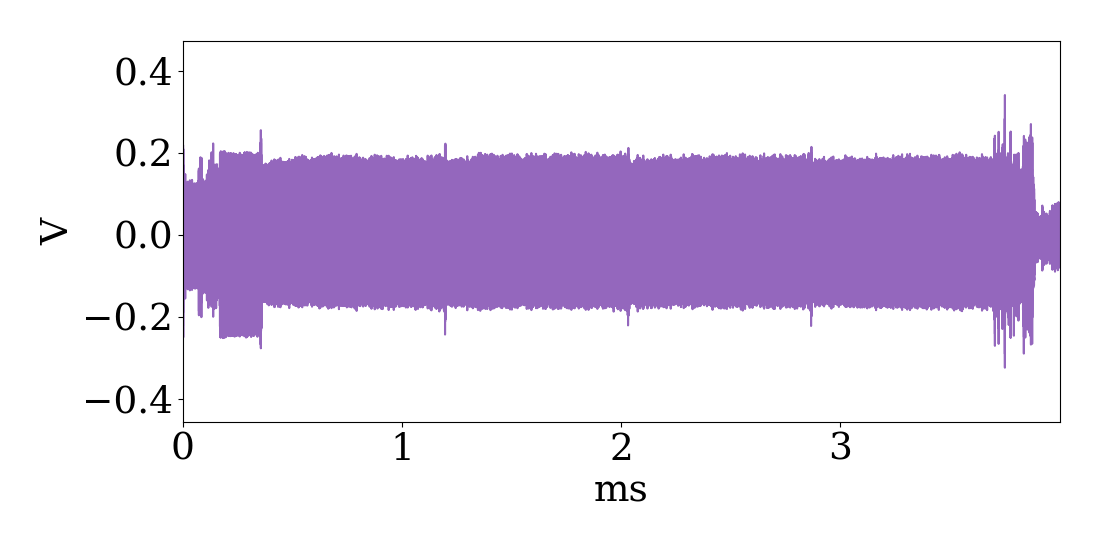}
    \caption{Dense layer, MNIST CNN, $N_e=16$}
    \label{fig_MNIST-CNN_dense_1}
\end{subfigure}\hspace{\fill}
\begin{subfigure}[t]{0.49\textwidth}
    \includegraphics[width=1\textwidth]{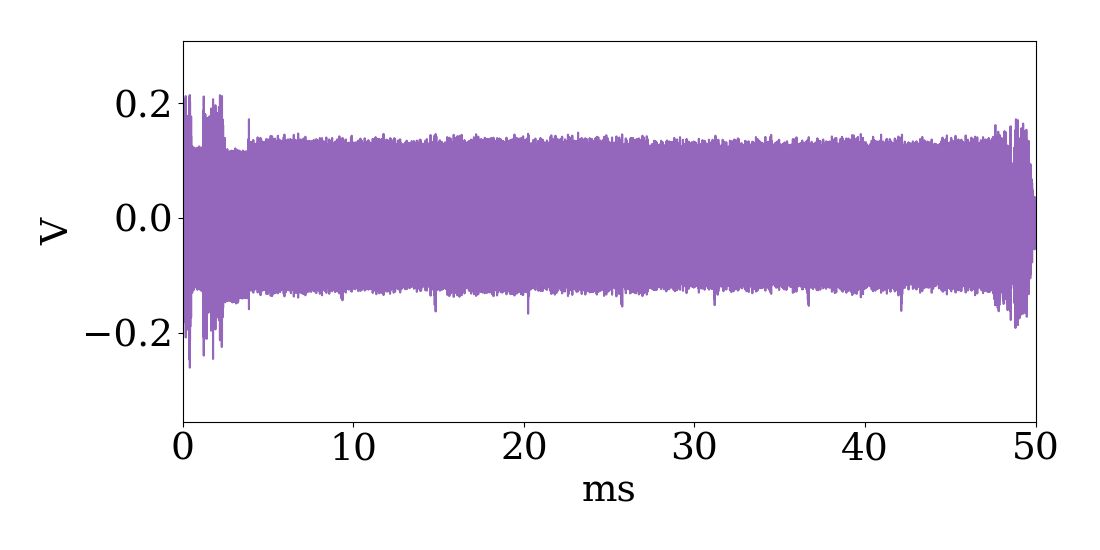}
    \caption{Dense layer, Cifar-10 CNN, $N_e=32$\vspace{5pt}}
    \label{fig_CIFAR-CNN_dense_1}
\end{subfigure}

\caption{Overview of dense layer EM activity corresponding to general cases}
\label{trace_number_neurons}
\end{figure}

\paragraph{Observations.}
We observe significant EM activity that results from neurons handling as illustrated in Fig.~\ref{trace_number_neurons}. Observed patterns are mainly composed of uniform blocks separated by clear spikes (especially visible on Fig.~\ref{fig_MNIST-MLP_dense_1}). Logically, related EM pattern length directly depends on the number of inputs to the layer and therefore to neurons. Then, dense layers managing broader inputs are easier to analyse. By counting the $N_p$ spike-separated blocks of each dense layers of our MLP and CNN, we checked the link between this number of occurrences $N_p$ and the number of neurons $N_e$ of the layer. Fig.~\ref{fig_MNIST-MLP_dense_1} and \ref{fig_MNIST-MLP_dense_2} illustrate the two dense layers of the MNIST MLP model with respectively 8 and 4 patterns corresponding to $N_e=32$ and $N_e=16$ neurons. As well, Fig~\ref{fig_MNIST-CNN_dense_1} and~\ref{fig_CIFAR-CNN_dense_1} show dense layers with respectively $N_e=16$ (4 patterns) and $N_e=32$ (8 patterns) neurons.

\paragraph{Limitations (special cases).}
As neurons are grouped in sets of 4, some special cases occur when remaining neurons still need to be computed. The line 10 of Alg.~\ref{alg_dense} checks remainders and handles them throughout the following \texttt{for} loop. Neurons are then handled one by one as only a single sum is initialised then used in \texttt{mac} call at line 15. To illustrate this phenomenon, we trained an additional MLP model on MNIST (noted SP-MLP), composed of 4 dense layers with respectively 23, 18, 13 and 10 neurons. These correspond to different remainders.

\begin{figure}[h!]
\centering
\begin{subfigure}[t]{0.99\textwidth}
    \centering
    \includegraphics[width=1\textwidth]{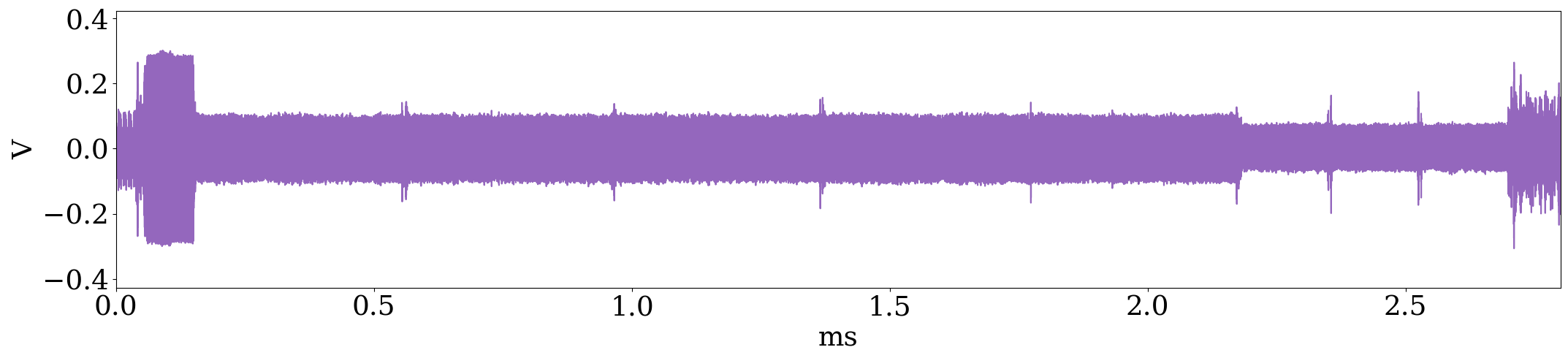}
    \caption{$1^{st}$ dense, \textit{$N=23$ $\Leftrightarrow 5+3$ patterns}\vspace{10pt}}
    \label{23_neur_MLP}
\end{subfigure}

\begin{subfigure}[t]{0.49\textwidth}
    \includegraphics[width=1\textwidth]{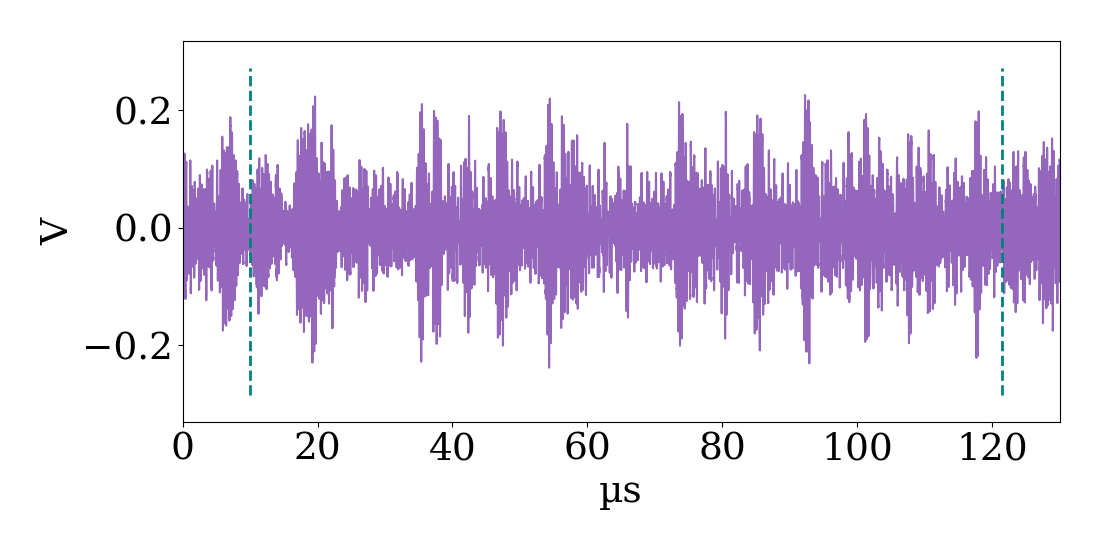}
    \caption{$2^{nd}$ dense \textit{ $N=18$ $\Leftrightarrow 4+2$ patterns}\vspace{10pt}}
    \label{18_neur_MLP}
\end{subfigure}\hspace{\fill}
\begin{subfigure}[t]{0.49\textwidth}
    \includegraphics[width=1\textwidth]{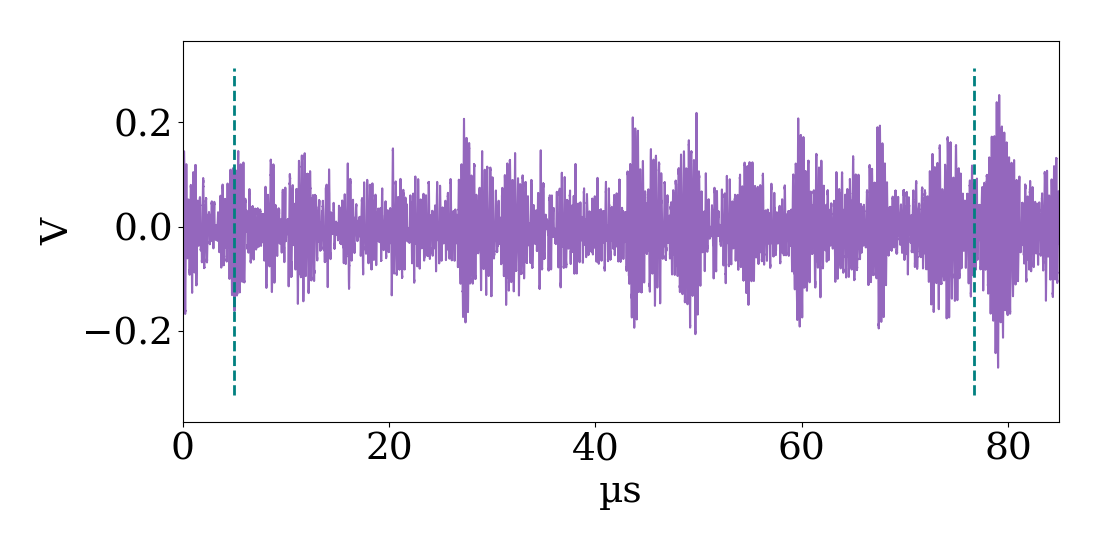}
    \caption{$3^{rd}$ dense \textit{ $N=13$ $\Leftrightarrow 3+1$ patterns}}
    \label{13_neur_MLP}
\end{subfigure}

\begin{subfigure}[t]{0.49\textwidth}
    \includegraphics[width=1\textwidth]{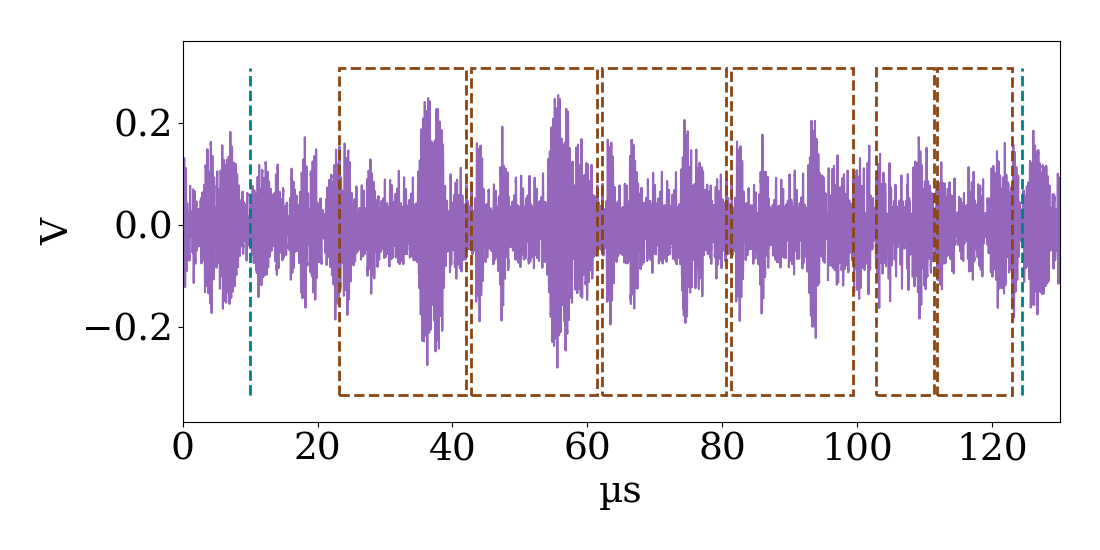}
    \caption{Same as (b) \textbf{with triggers}}
    \label{framed_18_neur_MLP}
\end{subfigure}\hspace{\fill}
\begin{subfigure}[t]{0.49\textwidth}
    \includegraphics[width=1\textwidth]{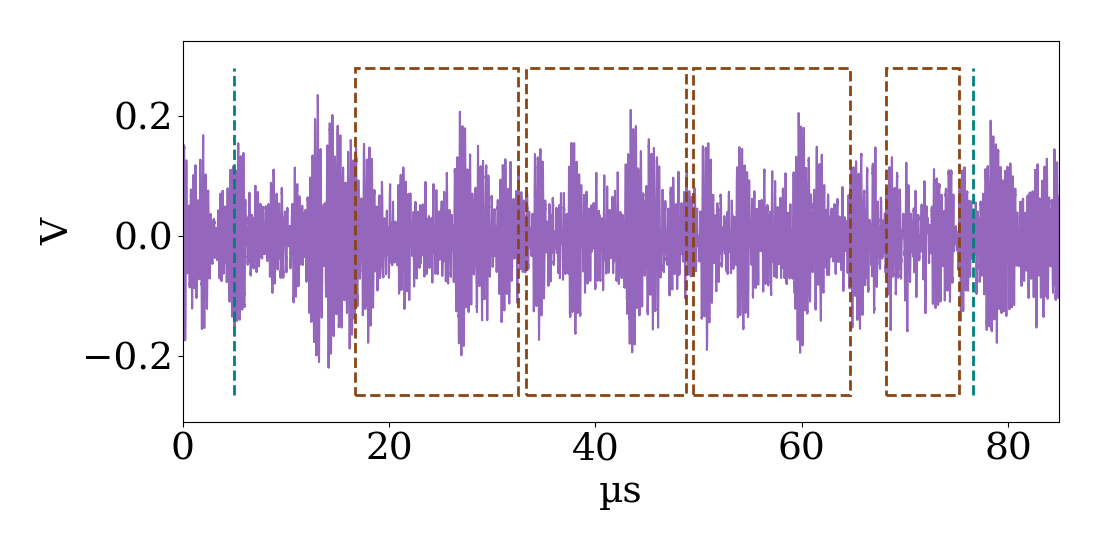}
    \caption{Same as (c) \textbf{with triggers}\vspace{5pt}}
    \label{framed_13_neur_MLP}
\end{subfigure}

\caption{EM activity overview of dense layers for the custom SP-MLP model. Expected patterns clearly appear for the first layer, not for the $2^{nd}$ and $3^{rd}$ 
}
\label{trace_number_neurons_sp}
\end{figure}

Fig.~\ref{trace_number_neurons_sp} shows EM activity of each layer for SP-MLP. For the 23-neuron layer, we clearly observe on Fig.~\ref{23_neur_MLP} 3 blocks that stand out after the core sequence of 5 patterns. This directly matches what is expected with 5 groups of 4 neurons completed with the 3 remaining ones managed independently. However, similar analysis cannot be performed on the two other traces. This difficulty comes from input tensor shape reduction from first layer to the second and third ones. To verify that neurons are managed in the same way for both of these, triggers are raised and lowered inside the outer \texttt{for} loop over \texttt{rowCnt} (represented as rectangles in Fig.~\ref{framed_18_neur_MLP} and~\ref{framed_13_neur_MLP}). In addition, it illustrates problematic pattern shape changes induced by usage of triggers around observed piece of code. Indeed, they become more visible in Fig.~\ref{framed_18_neur_MLP} and~\ref{framed_13_neur_MLP} compared to Fig.~\ref{18_neur_MLP} and~\ref{13_neur_MLP}.

\label{exp_section:dense}

\subsection{Activation layer}
\label{exp_section:act_func}
\begin{figure}[t!]
\centering

\begin{subfigure}[h]{0.49\textwidth}
    \includegraphics[width=1\textwidth]{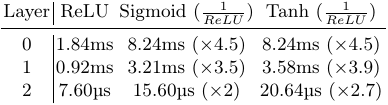}
    \vspace*{0.2mm}
    \caption{Layer duration according to AF}
    \label{tab:AF_duration}
\end{subfigure}\hspace{\fill}
\begin{subfigure}[h]{0.49\textwidth}
    \includegraphics[width=1\textwidth]{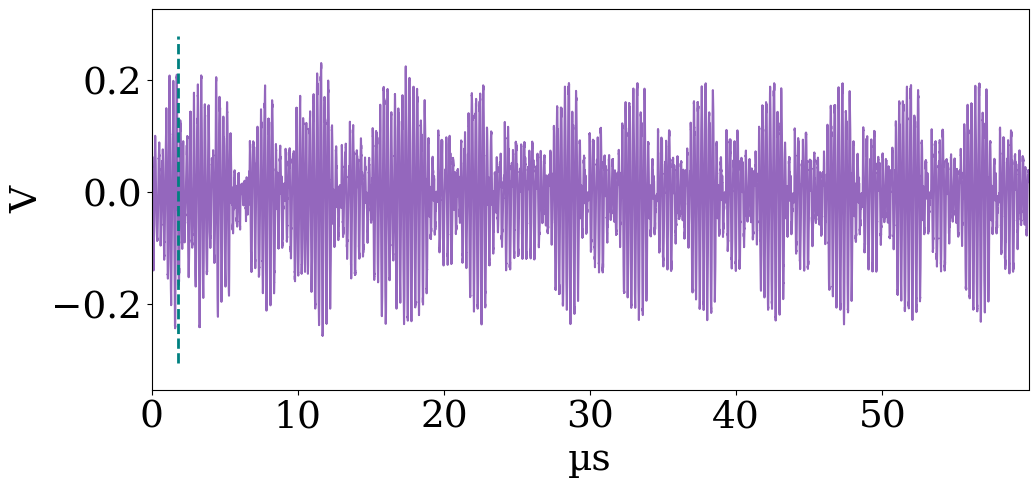}
    \caption{$2^{nd}$ ReLU layer\vspace{10pt}}
    \label{ReLU_AF_1}
\end{subfigure}
\vspace{10pt}
\begin{subfigure}[t]{0.49\textwidth}
    \includegraphics[width=1\textwidth]{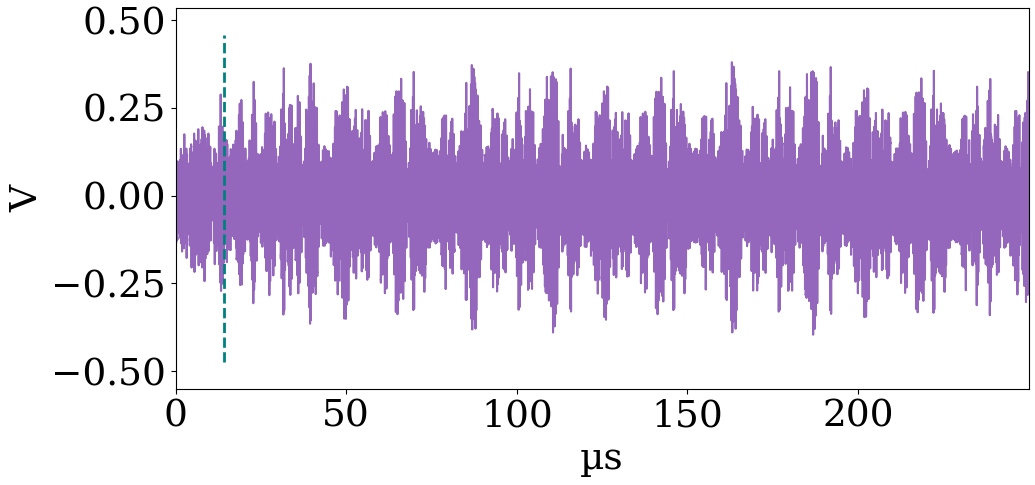}
    \caption{$2^{nd}$ Sigmoid layer}
    \label{Sigmoid_AF_1}
\end{subfigure}\hspace{\fill}
\begin{subfigure}[t]{0.49\textwidth}
    \includegraphics[width=1\textwidth]{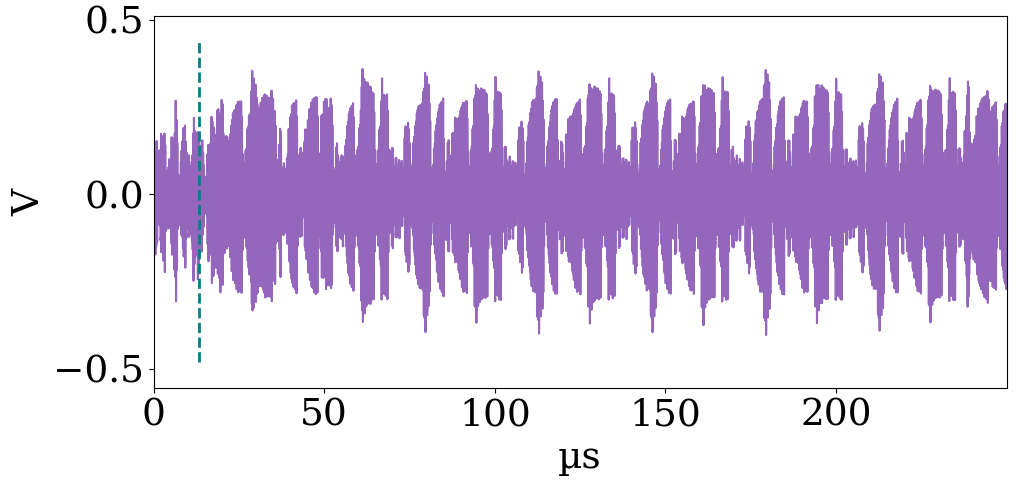}
    \caption{$2^{nd}$ Tanh layer}
    \label{Tanh_AF_1}
\end{subfigure}

\caption{Zoom over the beginning of second activation layers with their duration}
\label{AF_patterns}
\end{figure}

\noindent The nature of the Activation Functions (AF) is an important information since it strongly affects input flow from one layer to the next. It is especially useful for parameter extraction as it can modify layers input distribution and potentially their signs (e.g., $\forall x \in \mathbb{R}, ReLU(x)\geq0$).  
Many different AF exist with Sigmoid, Tanh, Softmax (mainly for the model output normalization) and ReLU as the most popular.  
The first three imply an exponential or division computation that are time consuming. ReLU is predominantly the most popular AF because of training efficiency and its low computation requirements. Therefore, we mainly focus on the distinction between ReLU from Sigmoid and Tanh functions (i.e., the extraction goal is to answer the question: \textit{is the AF ReLU or not?}). Measuring computation time of entire AF layer can give strong evidence (as also investigated in~\cite{batina2019csi}) because ReLU function is processed faster than the other two as shown in Table (a) from Fig.~\ref{AF_patterns}. Such approach takes on its full meaning when considering potential template capacity of the attacker. 

However, we observe that looking at the EM patterns gives additional hints that help distinguish ReLU. For that purpose, we trained three MNIST CNN models with the same architecture but different AF (one per model: ReLU, Sigmoid or Tanh).
Traces of Fig.~\ref{AF_patterns} are zoomed in on the beginning of each second AF layer. Trace~\ref{ReLU_AF_1}, corresponding to ReLU, exhibits regular and distinguishable patterns (duration of few $\mu s$). Sigmoid and Tanh traces also present groups of peaks repeated throughout traces. However, these are more complex and less explicit than ones related to ReLU and the order of magnitude of their duration is greater, especially for Sigmoid traces. From these observations, distinction between ReLU and the two other AF is quite straightforward. Noteworthy, it is not a clear-cut between Sigmoid and Tanh.


\section{Architecture extraction methodology}
\label{exp_section:overall}
At first, an adversary may exploit the nature of the task performed by the model. In some cases, it can give hints about the type of the victim model. A model performing a computer vision task is likely to be built around CNN principle. However, there are no strict rules that match a task to a model type. We claim that task knowledge may provide a set of possible layer types that a first analysis of the trace overall structure must confirm. Moreover, preliminary knowledge also encompasses classical deep learning practices, more precisely the \textit{logical} order of layers. If the first layer is identified as a convolutional one, a standard association is another conv. layer or a pooling one. 

The first information to extract is the overall structure of $\mathcal{A}_M$ with two information: the number of layers, $L$ and their natures. Then, the attacker focuses on each layer one after the other and, according to their nature, extracts a set of hyper-parameters easing to design a substitute model $M'_{\Theta}$. 

\begin{figure}[t!]
\centering
\begin{subfigure}[t]{0.49\textwidth}
    \includegraphics[width=1\textwidth]{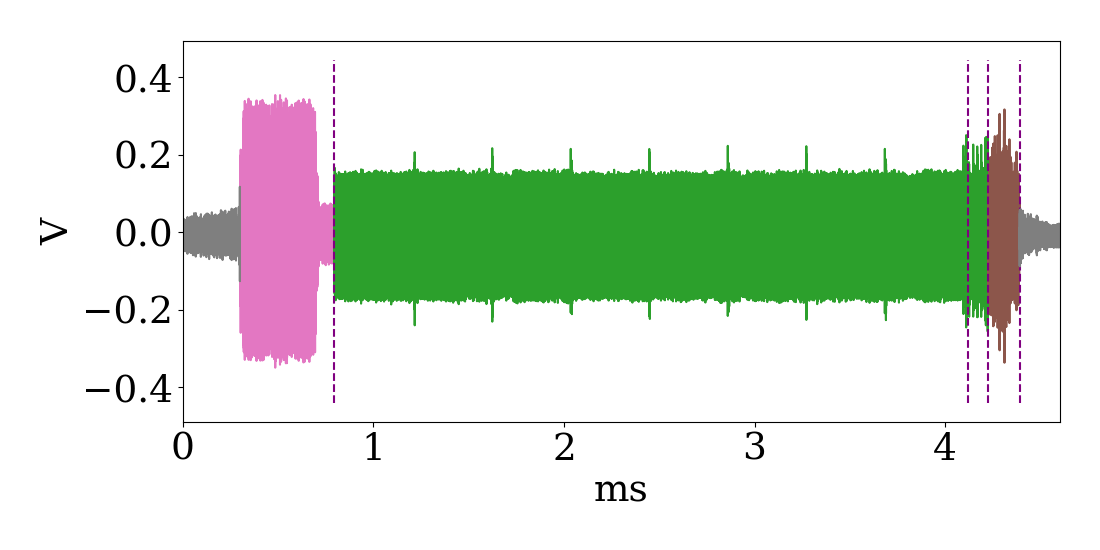}
    \caption{MLP colored layers}
    \label{MLP_inf_colored}
\end{subfigure}\hspace{\fill}
\begin{subfigure}[t]{0.49\textwidth}
    \includegraphics[width=1\textwidth]{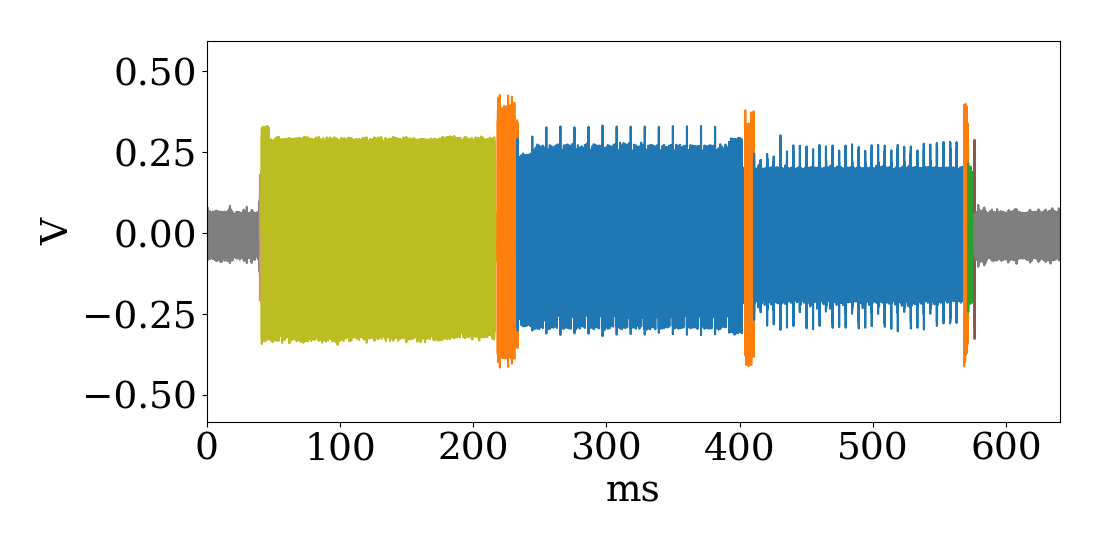}
    \caption{Cifar-10 CNN colored layers\vspace{10pt}}
    \label{cifar10_CNN_inf_colored}
\end{subfigure}

\begin{subfigure}[t]{0.49\textwidth}
    \includegraphics[width=1\textwidth]{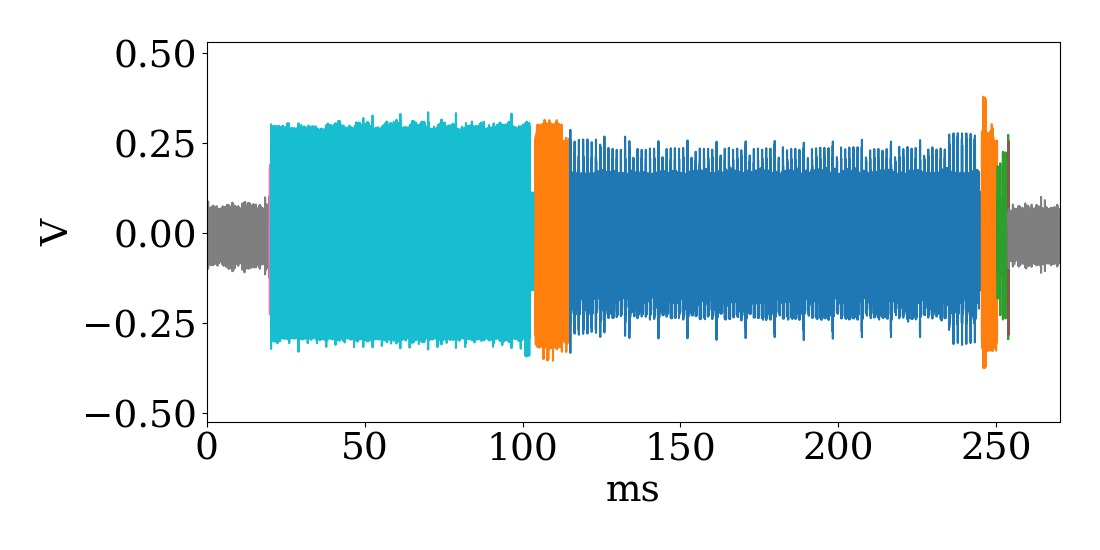}
    \caption{MNIST CNN colored layers}
    \label{mnist_CNN_inf_colored}
\end{subfigure}\hspace{\fill}
\begin{subfigure}[t]{0.49\textwidth}
    \includegraphics[width=1\textwidth]{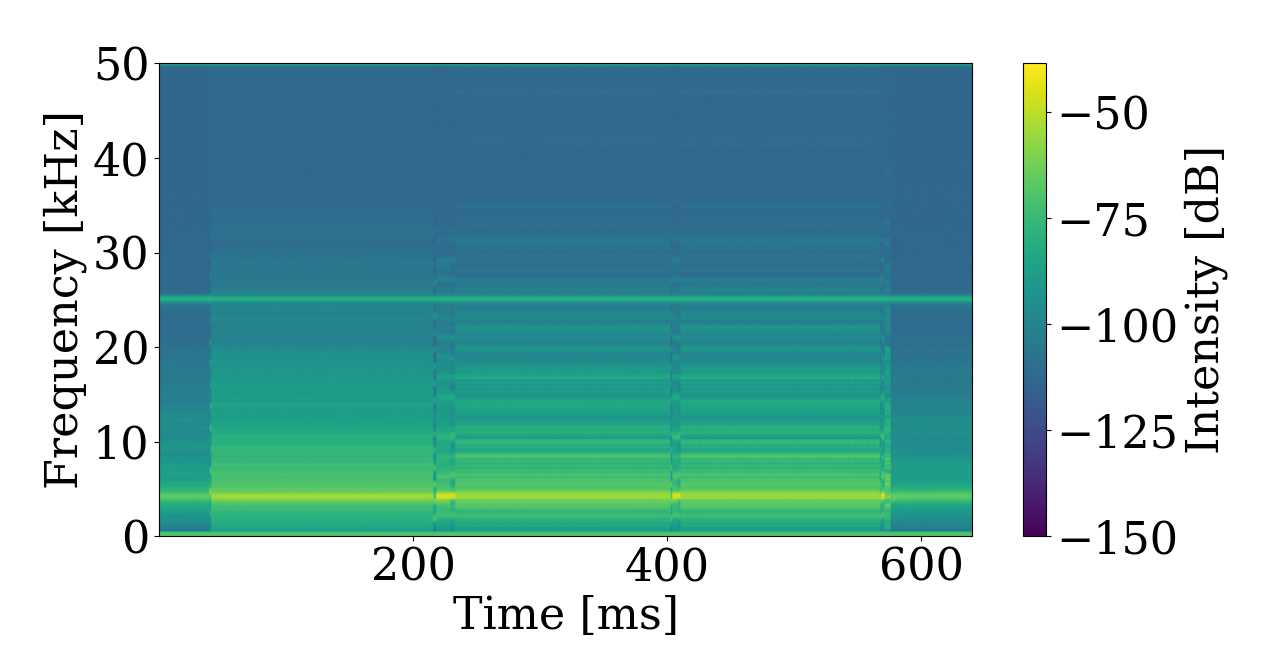}
    \caption{Cifar-10 CNN spectrogram\vspace{5pt}}
    \label{cifar10_CNN_inf_spectro}
\end{subfigure}

\caption{Average EM trace for the 3 models. Spectrogram for Cifar-10 CNN.}
\label{trace_entire_inf}
\end{figure}

\paragraph{Step 1: Finding the number of layers.}
When analyzing the general shape of the averaged EM trace, the very first observation is that it can be straightforwardly split into several blocks by identifying the boundaries that separate them. A frequency spectrum can highlight the separation between the blocks. The Fig.~\ref{cifar10_CNN_inf_spectro} illustrates how this processing can made easier the layer splitting for the Cifar-10 CNN. Without difficulty, for the three models studied in this work, this simple analysis gives the exact number $L$ of layers for each model architecture $\mathcal{A}_M$. Fig.~\ref{trace_entire_inf} illustrates the identified blocks for our models with the same colors as in Fig.~\ref{diagram_models_archi}. To validate that layers truly correspond to blocks and for illustration purpose only, triggers have been added to clearly mark out layer boundaries. 

\paragraph{Step 2: Identifying layers' nature.}
Having the position of the $L$ layers on the EM trace, the attacker knows the intrinsic layer complexity that directly influences the execution times. Typically, pooling and AF layers are far less complex than the main computational ones (in our case, dense and conv.) that they usually follow. Thus, the real challenge is to correctly initiate the extraction process by identifying the first layer as a dense or a conv. layer. A correct extraction of the first layer is fundamental, since the \textit{course} of the attack, as well as assumptions made on $\mathcal{A}_M$, come out of this \textit{good start}. To successfully achieve this distinction between dense and conv. layer, the attacker relies on a twofold analysis: 
\begin{itemize}
    \item \textit{Complexity / execution times.} First insight is provided by simple timing analysis. Although insufficient, this analysis may dispel doubts between some layer hypotheses. Complexity is estimated in different ways in the literature but a standard way is to consider the number of Multiplications and Accumulations (MACs). Such metric directly stems from computations and input/output dimensions. MAC complexity of conv. and dense layers are as follow: $MAC_{Conv2D} = (Z^2 \times C_{in}) \times (H_{out}^2 \times C_{out})$, $MAC_{dense} = H_{in} \times H_{out}$. 

Complexity of the first layer of both MNIST models are 112896 MACs for the CNN and 25088 MACs (x4.5 less) for the MLP. Thus, usually, conv. layer takes longer to execute than dense one (with similar inputs)\footnote{Obviously, if the dense layer had x4.5 more neurons, MAC complexity between dense and conv. would be equal but having a too large number of neurons (i.e. trainable parameters) for dense layers is usually unsuitable with classical overfitting issues.}.
    \item \textit{Patterns:} the attacker mostly leverages on specific EM activities. As detailed in Section~\ref{exp_section}, the regularities and repetitions of patterns are very different between dense and conv. layers and confusion is very unlikely. Thus, layer's nature is identified before accurately extracted the related hyper-parameters. 
\end{itemize}

Once the first layer is identified, ML expertise and usual layers sequence knowledge can provide additional hints about next layers nature. 
This encourages the attacker to extract $\mathcal{A}_M$ by analysing model layers consecutively.

\paragraph{Step 3: Extracting hyper-parameters.}
Once the overall architecture is known, the attacker follows the analysis presented in Section~\ref{exp_section} to recover hyper-parameters.

\section{Discussions and Perspectives}
\label{discu}
Our main objective is to estimate how much information about the architecture of a victim model an adversary can extract by exploiting limited side-channel traces. This information can considerably help the attacker to perform powerful adversarial attacks against the integrity or confidentiality of the victim model. Even though we demonstrate that very critical information can be deduced by the methodical analysis of both an EM trace and the implementation details of the deployment library (here, CMSIS-NN), the use of pattern extraction and recognition tools may significantly ease extraction process by automating most steps and potentially help in extracting more hyper-parameters. 
We believe that this work paves the way for such further analysis as other efforts that would aim to widen the scope of analyzed architectures and layers. More particularly, batch normalization or tensor arithmetic layers as in state-of-the-art ResNet models or Attention blocks as in Transformer models are relevant candidates for future works. We also highlight that variants of the standard convolution have been proposed for (computational) efficiency purpose (e.g., Depthwise Separable Convolution)~\cite{nguyen2022evaluation}. To the best of our knowledge, all these variants rely on highly repetitive and regular computations as the convolution with im2col and GeMM, therefore it should not fundamentally differ from what we exposed in this work.

Moreover, the impact of experimental setup simplifications (i.e., disable of interruptions and cache optimizations) must be studied in complementary studies. Such changes could harden hyper-parameters recovery and directed research focused on protection dedicated to model architecture. To the best of our knowledge, they are very few of them, especially for microcontroller platform. Authors of \cite{luo2022nnrearch} propose an obfuscation strategy for FPGA accelerators. They leverage on optimization parameters of convolution computation to mitigate EM emanations coming from conv. layers. Porting similar protection to microcontroller could be challenging due to limited resources and model performance to be preserved.

\section{Conclusion}
\label{conclu}
When dealing with DNN model extraction attack, architecture is crucial. Whatever the attacker's objective, a complete or even partial recovery provides an essential advantage. With this work, we highlight that the attack surface for such a threat is significantly extended by side-channel analysis. More importantly, regarding our application scope, we demonstrate that there is no need for complex exploitation methods (e.g., with heavy supervised profiling step) because of the strong repetitiveness and regularity of most of performed computations that make SEMA a surprisingly powerful tool. Typically, the Russian dolls effect that we exploit for conv. layers (enabling the recovery, one after the other, of several important hyper-parameters) is highly representative of this confidentiality flaw that we claim to be a very worrying concern. 

Although we highlight some limitations or more complex special cases that need to be handled in future works, our concern is not based only on the relative simplicity of the attack, but also on the hard challenges related to the development of efficient \textit{and} practical protections, compliant with the constrains of 32-bit microcontrollers. With ongoing regulatory frameworks for AI systems and upcoming security certification actions, model architecture obfuscation appears as the key defense challenge, that we urgently need to solve in order to bring robustness to the large-scale deployment of ML systems.

\section*{Acknowledgements}
\noindent This work is supported by (CEA-Leti) the EU project InSecTT (ECSEL JU 876038) 
and by ANR (Fr) in the framework \textit{Investissements d’avenir} program (ANR-10-AIRT-05, irtnanoelec);  and (Mines Saint-Etienne) by ANR PICTURE program (AAPG2020). This work benefited from the French Jean Zay supercomputer with the AI dynamic access program.

\bibliographystyle{splncs04}
\bibliography{biblio/biblio}

\end{document}